\documentclass[final,1p,sort&compress,times]{elsarticle}
\usepackage{amsmath,amssymb,amscd}
\newcommand{\al}{\alpha}
\newcommand{\be}{\beta}
\newcommand{\de}{\delta}

\newcommand{\vep}{\varepsilon}
\newcommand{\ga}{\gamma}
\newcommand{\ka}{\kappa}
\newcommand{\la}{\lambda}

\newcommand{\si}{\sigma}

\newcommand{\vp}{\varphi}
\newcommand{\ze}{\zeta}
\newcommand{\De}{\Delta}
\newcommand{\Ga}{\Gamma}
\newcommand{\La}{\Lambda}
\newcommand{\Si}{\Sigma}
\newcommand{\bde}{\boldsymbol{\delta}}

\newcommand{\bk}{\mathbf{k}}

\newcommand{\bll}{\boldsymbol{\ell}}

\newcommand{\bn}{\mathbf{n}}

\newcommand{\bbs}{\mathbf{s}}

\newcommand{\bx}{\mathbf{x}}

\newcommand{\bsi}{{\boldsymbol{\si}}}
\newcommand{\bxi}{{\boldsymbol{\xi}}}

\newcommand{\CC}{{\mathbb C}}
\newcommand{\NN}{{\mathbb N}}
\newcommand{\RR}{{\mathbb R}}

\newcommand{\cB}{{\mathcal B}}

\newcommand{\cE}{{\mathcal E}}
\newcommand{\cF}{{\mathcal F}}
\newcommand{\cH}{{\mathcal H}}

\newcommand{\cP}{{\mathcal P}}

\newcommand{\cZ}{{\mathcal Z}}
\def\Esc{E^{\mathrm{sc}}}
\def\Hsc{H^{\mathrm{sc}}}
\def\Zsc{Z^{\mathrm{sc}}}
\newcommand{\pa}{\partial}

\def\ket#1{|#1\rangle}

\renewcommand{\le}{\leqslant}
\renewcommand{\ge}{\geqslant}
\def\bbuildrel#1_#2^#3{\mathrel{\mathop{\kern0pt #1}\limits_{#2}^{#3}}}
\newcommand{\tends}[1]{\bbuildrel{\hbox to 2em{\rightarrowfill}}_{#1}^{}}
\newcommand{\erf}{\operatorname{erf}}
\newcommand{\erfi}{\operatorname{erfi}}
\newcommand{\sgn}{\operatorname{sgn}}
\newcommand{\Li}{\operatorname{Li}}
\newcommand{\sech}{\operatorname{sech}}
\newcommand{\tr}{\operatorname{tr}}

\newcommand{\iu}{\mathrm i}
\newcommand{\e}{\mathrm e}
\newcommand{\Or}{\mathrm{O}}
\newcommand{\diff}{\mathrm{d}}

\newcommand{\im}{\operatorname{Im}}
\newcommand{\Bs}{B_{\mathrm s}}
\newcommand{\kB}{k_{\rm B}}

\newcommand{\Eg}{E_{\mathrm g}}
\newcommand{\half}{1\kern -.7pt/2}
\newcommand{\HS}{\mbox{for the HS chain}}
\newcommand{\PF}{\mbox{for the PF chain}}
\newcommand{\FI}{\mbox{for the FI chain}}
\let\case\tfrac
\begin{document}
\begin{frontmatter}
  \title{Thermodynamics of
  spin chains of Haldane--Shastry type and one-dimensional vertex models}
\author[AE]{Alberto Enciso}
\author[UCM]{Federico Finkel}
\author[UCM]{Artemio Gonz\'alez-L\'opez\corref{cor}}

\cortext[cor]{Corresponding author}

\address[AE]{Instituto de Ciencias
   Matem\'aticas, Consejo Superior de Investigaciones Cient\'\i ficas, 28049 Madrid, Spain}
\address[UCM]{Departamento de F\'\i sica Te\'orica II, Universidad Complutense de Madrid, 28040 Madrid,
  Spain}
\date{April 13, 2012}
\begin{abstract}
  We study the thermodynamic properties of spin chains of Haldane--Shastry type associated with
  the $A_{N-1}$ root system in the presence of a uniform external magnetic field. To this end, we
  exactly compute the partition function of these models for an arbitrary finite number of spins.
  We then show that these chains are equivalent to a suitable inhomogeneous classical Ising model
  in a spatially dependent magnetic field, generalizing the results of Basu-Mallick et al.~for the
  zero magnetic field case. Using the standard transfer matrix approach, we are able to compute in
  closed form the free energy per site in the thermodynamic limit. We perform a detailed analysis
  of the chains' thermodynamics in a unified way, with special emphasis on the zero field and zero
  temperature limits. Finally, we provide a novel interpretation of the thermodynamic quantities
  of spin chains of Haldane--Shastry type as weighted averages of the analogous quantities over an
  ensemble of classical Ising models.
\end{abstract}
\end{frontmatter}
\numberwithin{equation}{section}
\section{Introduction}\label{sec.intro}

In this paper we study a class of $\mathrm{su}(m)$ spin chains whose Hamiltonian can be
collectively written as
\begin{equation}\label{cH}
\cH=\sum_{1\le i<j\le N}J_{ij}(1-S_{ij})-\sum_{\al=1}^{m-1}B_\al S^\al\,,
\end{equation}
where the $B_{\al}$'s are real constants and the interactions $J_{ij}$ are chosen as described
below (see Eqs.~(1.7)). In the previous formula, the operators $S_{ij}$ act on a state
\[
\ket{s_1,\dots,s_N}\,,\qquad s_i\in\{1,\dots,m\}\,,
\]
of the canonical spin basis by permuting the $i$-th and $j$-th spins:
\[
S_{ij}\ket{s_1,\dots,s_i,\dots,s_j,\dots,s_N}=\ket{s_1,\dots,s_j,\dots,s_i,\dots,s_N}\,.
\]
The permutation operators $S_{ij}$ can be expressed in terms of the
(Hermitian) $\mathrm{su}(m)$ spin operators $T_k^\al$ with the normalization
$\tr(T^\al_kT^\be_k)=2\de^{\al\be}$ (where the subindex $k$ labels the chain sites) as~\cite{Po06}
\begin{equation}\label{Sij}
S_{ij}=\frac1m+\frac12\sum_{\al=1}^{m^2-1}T^\al_iT^\al_j\,.
\end{equation}
Let $S^\al_k$ denote the operator whose action on the canonical spin basis is given by
\begin{equation}\label{Skal}
S_k^\al\ket{s_1,\dots,s_N}=(\de^\al_{s_k}-\de^m_{s_k})
\ket{s_1,\dots,s_N}\,,\qquad \al=1,\dots,m-1\,,
\end{equation}
so that the operators $\iu S^\al_k$ span the standard Cartan subalgebra of $\mathrm{su}(m)$. The
operators $S^\al$ are then defined by
\begin{equation}
  \label{Sal}
  S^\al=\sum_{i=1}^NS^\al_i\,,\qquad \al=1,\dots,m-1\,.
\end{equation}
Thus the second sum in Eq.~\eqref{cH} can be interpreted as arising from the interaction with a
uniform external $\mathrm{su}(m)$ ``magnetic'' field\footnote{%
  The most general $\mathrm{su}(m)$ magnetic field term is of the form $T=\sum_{i=1}^NT_i$, where
\[
T_i\equiv\sum_{\al=1}^{m^2-1}b_\al T^\al_i
\]
is a traceless Hermitian matrix acting on the internal space of the $i$-th spin. By performing a
rotation in this internal space we can diagonalize the matrix $T_i$, effectively replacing it by a
traceless diagonal matrix. The latter matrix can in turn be expressed in the form
$\sum_{\al=1}^{m-1}B_\al S^\al_i$, which yields the last term in Eq.~\eqref{cH}.} of strengths
$(B_1,\dots,B_{m-1})$. Note that in the case $m=2$ (i.e., for spin $\half$) we can take
$\{T^1_k,T^2_k,T^3_k\}=\{\si^x_k,\si^y_k,\si^z_k\}$ and $S_k^1=\si^z_k$, where $\si^\nu_k$ is a
Pauli matrix acting on the $k$-th spin's Hilbert space. Hence Eq.~\eqref{Sij} adopts the more
familiar form
\begin{equation}
  \label{SijSalsu2}
  S_{ij}=\case12(1+\bsi_i\cdot\bsi_j)\,,
\end{equation}
and the Hamiltonian~\eqref{cH} reduces to
\begin{equation}\label{cH2}
  \cH=\sum_{1\le i<j\le N}\frac{J_{ij}}2\,\big(1-\bsi_i\cdot\bsi_j\big)-B\sum_{i=1}^N\si_i^z\,,
\end{equation}
with $B\equiv B_1$. In particular, the last term represents the interaction with a uniform
magnetic field parallel to the $z$ axis with strength (proportional to) $B$.

The three models we shall study are defined by the following choice of the interaction strengths
$J_{ij}$:

\medskip\noindent
$\bullet$ The \emph{Haldane--Shastry} (HS) chain~\cite{Ha88,Sh88}:
\begin{subequations}
\label{Jijs}
\begin{equation}
  \label{HSdef}
  J_{ij}=\frac{J}{2\sin^2(\xi_i-\xi_j)}\,,\qquad \xi_k=\frac{k\pi}N\,.
\end{equation}

\medskip\noindent
$\bullet$ The \emph{Polychronakos--Frahm} (PF) chain~\cite{Po93,Fr93}:

\begin{equation}
  \label{PFdef}
  J_{ij}=\frac{J}{(\xi_i-\xi_j)^2}\,,
\end{equation}
where $\xi_k$ is the $k$-th root of the Hermite polynomial of degree $N$.

\medskip\noindent
$\bullet$ The \emph{Frahm--Inozemtsev} (FI) chain~\cite{FI94}:

\begin{equation}
  \label{FIdef}
  J_{ij}=\frac{J}{2\sinh^2(\xi_i-\xi_j)}\,,
\end{equation}
\end{subequations}
where $\e^{2\xi_k}$ is the $k$-th root of the generalized Laguerre polynomial
$L_N^{\al-1}$ with $\al>0$. In all three cases, $J$ is a real constant whose sign determines the
model's ferromagnetic ($J>0$) or antiferromagnetic ($J<0$) character. Note that, while the sites
of the HS chain are equispaced\footnote{In fact, if $2\xi_k$ is interpreted as an angular
  coordinate, then the HS chain describes an array of spins equispaced on the unit circle, with
  long-range pairwise interactions inversely proportional to the square of the chord distance
  between the spins.}, this is not the case for the PF or FI chains.\medskip

We shall denote by
\begin{equation}\label{cH0}
\cH_0=\sum_{1\le i<j\le N}J_{ij}(1-S_{ij})
\end{equation}
the Hamiltonian of the chains~\eqref{cH} in the absence of a magnetic field. Following standard
terminology, we shall collectively refer to the chains~\eqref{Jijs}-\eqref{cH0} as spin chains of {\em
  Haldane--Shastry type}. They are all associated with the root system $A_{N-1}$, in the sense
that the interactions $J_{ij}$ depend only on the differences of the site coordinates $\xi_k$.
Although several generalizations of these chains to the $BC_N$ and $D_N$ root systems have been
considered in the literature \cite{YT96,EFGR05,BFGR08,BFG09,BFG11}, in this paper we shall
restrict ourselves to the above $A_{N-1}$-type models.

Spin chains of HS type are the simplest models in condensed matter physics exhibiting fractional
statistics~\cite{Ha91b}. Historically, the HS chain~\eqref{HSdef}-\eqref{cH0} was introduced as a
simplified version of the one-dimensional Hubbard model with long-range hopping, from which it can
be obtained in the half-filling regime when the on-site interaction tends to infinity~\cite{GR92}.
Soon after its introduction, it was shown that this chain is completely integrable, in the sense
that one can explicitly construct $N-1$ mutually commuting integrals of motion~\cite{FM93,BGHP93}.
As first observed by Polychronakos~\cite{Po93}, these integrals of motion can be obtained from those
of the {\em dynamical} spin Sutherland model~\cite{HH92} by means of the so-called ``freezing
trick''. In fact, the freezing trick can also be applied to derive the PF and FI chains from the
Calogero~\cite{MP93} and Inozemtsev~\cite{In96} spin dynamical models. In particular, these two
chains are also completely integrable. Apart from their integrable character, spin chains of HS
type appear in many areas of current interest in both physics and mathematics, such as quantum
chaos~\cite{BFGR08epl,BFGR09power}, supersymmetry~\cite{BUW99,BB06}, conformal field
theory~\cite{Ha91,BBS08,CS10}, the AdS-CFT correspondence~\cite{BKS03,BBL09}, one-dimensional
anyons~\cite{Gr09} and Yangian quantum groups~\cite{BGHP93,Hi95npb,Ba99,BE08}.

The partition functions of all three chains of HS type in the absence of a magnetic field, which
have been computed in closed form using again the freezing trick~\cite{Po94,FG05,BFGR10}, can be
expressed in a unified way by the formula
\begin{equation}
  \label{cZ0}
  \cZ_0(q) = \sum_{\bk\in\cP_N}\prod_{i=1}^r{\textstyle\binom{m+k_i-1}{k_i}}\cdot
  q^{\sum_{i=1}^{r-1}J\cE(K_i)}
  \prod_{i=1}^{N-r} (1-q^{J\cE(K'_i)})\,.
\end{equation}
Here $q\equiv\e^{-1/(k_{\mathrm B}T)}$, $\bk\equiv (k_1,\dots,k_r)$ is an element of the set
$\cP_N$ of partitions of $N$ with order taken into account, and the numbers
$K_i,K'_j\in\{1,\dots,N-1\}$ in Eq.~\eqref{cZ0} are positive integers defined by
\begin{equation}\label{KKp}
K_i=\sum_{j=1}^ik_j,\qquad
\{K'_1,\dots,K'_{N-r}\}=\{1,\dots,N-1\}\setminus\{K_1,\dots,K_{r-1}\}\,.
\end{equation}
Remarkably, the partition function $\cZ_0$ of the chains~\eqref{Jijs}-\eqref{cH0} depends on the chain under
consideration only through its \emph{dispersion relation}
\begin{equation}\label{cE}
\cE(i)=
\begin{cases}
  i(N-i)\,,& \HS\\
  i\,,& \PF\\
  i(\al+i-1)\,,\quad & \FI.
\end{cases}
\end{equation}

Using the previous expression for the partition function, Basu-Mallick et al.~\cite{BBH10} showed
that the spectrum of the spin chains of HS type is given by
\begin{equation}\label{energies}
E_0(\bn)=J\sum_{i=1}^{N-1}\de(n_i,n_{i+1})\,\cE(i)\,,\qquad \bn\equiv(n_1,\dots,n_N)\,,
\end{equation}
where
\begin{equation}\label{dejk}
\de(j,k)=\begin{cases}
  1\,, \quad & j>k\\
  0\,,& j\le k\,,
\end{cases}
\end{equation}
and the quantum numbers $n_i$ independently take the values $1,\dots,m$. The vectors
$\bde(\bn)\in\{0,1\}^{N-1}$ with components $\de_i(\bn)=\de(n_i,n_{i+1})$ are in fact the
celebrated {\em motifs} introduced by Haldane et al.~\cite{HHTBP92}. As pointed out in
Ref.~\cite{BBH10}, Eq.~\eqref{energies} defines a classical {\em inhomogeneous} one-dimensional
vertex model with bonds $n_1,\dots,n_N$ each taking $m$ possible values, where the contribution of
the $k$-th vertex is given by\footnote{By Eq.~\eqref{energies}, the first and last vertices do not
  contribute to the energy.} $J\de(n_{k-1},n_k)\cE(k-1)$. We shall show that this connection
between spin chains of HS type and vertex models still holds in the presence of a nonzero
$\mathrm{su}(m)$ magnetic field. This is indeed the key result on which we shall base our unified
approach to the analysis of the thermodynamics of all three chains of HS type~\eqref{cH}.

The study of the thermodynamic properties of spin chains of HS type was initiated by Haldane
himself, who used the spinon description of the spectrum to derive an indirect expression for the
entropy of the spin $\half$ HS chain~\cite{Ha91}. An explicit formula for the free energy of the PF
chain in the absence of a magnetic field appeared shortly afterwards without proof in
Ref.~\cite{Fr93}. In a subsequent publication~\cite{FI94}, Frahm and Inozemtsev obtained an
analogous expression for the FI chain using the transfer matrix method, and also computed its
magnetization for arbitrary magnetic field.

In this paper we have several objectives that we shall now summarize. In the first place, we shall
compute in closed form the partition function of the HS-type chains~\eqref{cH} in the presence of
an arbitrary $\mathrm{su}(m)$ magnetic field $(B_1,\dots,B_{m-1})$. Secondly, we shall use the
expression for the partition function to establish the equivalence of the latter chains to a
suitable classical inhomogeneous vertex model. We shall then take advantage of this connection to
determine the equilibrium thermodynamics of the spin \half{} chains of HS type in a unified and
systematic way. Finally, we shall use the previous results to uncover a novel connection between
spin chains of HS type and the classical Ising model.

The paper is organized as follows. In Section~\ref{sec.pfs} we compute the chains' partition
functions in closed form by means of the freezing trick. Following the approach of
Ref.~\cite{BBH10}, in Section~\ref{sec.vms} we construct a generating function for the partition
function in terms of complete homogeneous symmetric polynomials. We then define a similar function
for the corresponding vertex models, and prove that both generating functions coincide for
arbitrary values of their arguments. This shows that the energies of each of the chains~\eqref{cH}
are identical to those of its associated vertex model. Exploiting this connection, in
Section~\ref{sec.thermo} we use the transfer matrix method to find explicit expressions for the
thermodynamic functions of the chains~\eqref{cH} with spin $\half$. In the case of the PF chain,
we derive an exact formula for the free energy in terms of the dilogarithm function in an
arbitrary magnetic field, and show that its magnetization can be expressed in terms of elementary
functions. Section~\ref{sec.zeroB} is devoted to a detailed study of the zero magnetic field case.
In particular, we show that the susceptibility can be expressed for all three chains of HS type in
terms of the error function, discuss the connection of these chains with two-level systems, and
derive low-temperature asymptotic expansions the chains' thermodynamic functions. In
Section~\ref{sec.zeroT} we study the zero temperature limit in the presence of an arbitrary
external magnetic field in both the ferromagnetic and antiferromagnetic regimes. We explicitly
show that in both regimes there is a phase transition, as was to be expected on general
grounds~\cite{Fr93}. In Section~\ref{sec.Ising} we present a novel interpretation of the
thermodynamic quantities of spin chains of HS type as weighted averages of the analogous
quantities over an ensemble of classical Ising chains. The paper ends with a short section where
we compare our results with previous related work, and comment on possible extensions thereof.

\section{Partition functions}\label{sec.pfs}

In this section we shall derive a closed form expression for the partition functions of the HS
chains~\eqref{cH} in the presence of a constant $\mathrm{su}(m)$ magnetic field. For definiteness,
we shall deal with the PF chain, whose interactions are defined by Eq.~\eqref{PFdef}. As we
mentioned above, this chain is obtained by applying the freezing trick to the $\mathrm{su}(m)$
spin Calogero model in a constant magnetic field, whose Hamiltonian we shall take as
\begin{equation}
  \label{H}
  H=H_0-\frac{2a}J\sum_{\al=1}^{m-1} B_\al S^\al\,,
\end{equation}
where $a>1/2$ and
\begin{equation}\label{H0}
H_0=-\sum_{i=1}^N \pa^2_{x_i}+a^2\sum_{i=1}^N x_i^2+\sum_{1\le i\ne j\le N}\frac{a(a-
S_{ij})}{(x_i-x_j)^2}\,.
\end{equation}
More precisely, let
\begin{equation}\label{Hsc}
\Hsc=-\sum_{i=1}^N \pa^2_{x_i}+a^2\sum_{i=1}^N x_i^2+\sum_{1\le i\ne j\le N}\frac{a(a-1)}{(x_i-x_j)^2}
\end{equation}
denote the Hamiltonian of the scalar Calogero model~\cite{Ca71}, and define
\begin{equation}\label{tcH}
\tilde\cH(\bx)=\sum_{1\le i<j\le N}\frac{1-S_{ij}}{(x_i-x_j)^2}-\frac1J\sum_{\al=1}^{m-1}B_\al S^\al\,,
\end{equation}
so that
\begin{equation}\label{HcH}
H=\Hsc+2a\,\tilde\cH({\bx})\,,\qquad
\cH=J\,\tilde\cH(\bxi)\,.
\end{equation}
It is well known~\cite{CS02} that the chain sites $\bxi$ of the PF chain are the coordinates of
the unique minimum of the scalar potential
\begin{equation}\label{U}
U(\bx)=\sum_{i=1}^N x_i^2+\sum_{1\le i\ne j\le N}\frac{1}{(x_i-x_j)^2}
\end{equation}
in the principal Weyl chamber of $A_{N-1}$ type $x_1<x_2<\cdots<x_N$, where the motion of the
particles of the Calogero model can be restricted due to the singularities of its Hamiltonian.
{F}rom this fact and Eq.~\eqref{H} it follows that in the large $a$ limit the eigenfunctions of
the spin Calogero model become sharply peaked around the sites of the PF chain, so that the
$\mathrm{su}(m)$ degrees of freedom effectively decouple from the dynamical ones. By
Eq.~\eqref{HcH}, in this limit the energies of the model~\eqref{H} are approximately given by the
formula
\begin{equation}\label{EEe}
E_{ij}\simeq \Esc_i+\frac{2a}J\,e_j\,,
\end{equation}  
where $E_i$ and $e_j$ denote two arbitrary eigenvalues of $\Hsc$ and $\cH$, respectively. Although
the spectra of both the scalar and the spin Calogero models can be easily computed, the previous
formula cannot be directly used to compute the spectrum of the PF chain. Indeed, it is not obvious
a priori which energies of $H$ and $\Hsc$ can be combined to yield an energy of the PF chain.
However, from Eq.~\eqref{EEe} it is immediate to obtain the following \emph{exact} relation between
the partition functions $Z$, $\Zsc$ and $\cZ$ of the Hamiltonians $H$, $\Hsc$ and $\cH$:
\begin{equation}
  \label{Zs}
  \cZ(T)=\lim_{a\to\infty}\frac{Z(2aT/J)}{\Zsc(2aT/J)}\,.
\end{equation}
This formula, first derived by Polychronakos~\cite{Po94}, is the mainstay of the freezing trick
method; see \cite{EFGR08jnmp,En09} for a rigorous proof.

We shall next evaluate both the numerator and the denominator in the previous equation. To begin
with, the partition function of the scalar Calogero model is easily computed~\cite{Po94}, with the
result
\begin{equation}
  \label{ZscPF}
  \Zsc(2aT/J)=q^{\frac{J\Eg}{2a}}\prod_{i=1}^N(1-q^{iJ})^{-1}\,,
\end{equation}
where
\begin{equation}
  \label{Esc0}
  \Eg=a^2N(N-1)+aN
\end{equation}
is the ground state energy.

Let us now turn to the partition function of $H$. Consider, to this end, the spin functions
\begin{equation}\label{psins}
  \psi_{\bll,\bbs}(\bx)=\rho(\bx)\La\Big(\,{\textstyle\prod\limits_{i=1}^N} x_i^{\ell_i}\,\ket{s_1,\dots,s_N}\Bigr)\,,
\end{equation}
where $\ell_i\in\NN\cup\{0\}$, $\bll=(\ell_1,\dots,\ell_N)$, $\bbs=(s_1,\dots,s_N)$,
\[
\rho(\bx)=\e^{-a r^2/2}\prod_{1\le i<j\le N}|x_i-x_j|^a\,,\qquad r^2\equiv\sum_{i=1}^Nx_i^2\,,
\]
is the ground state of the scalar Hamiltonian $\Hsc$ and $\La$ is the total symmetrizer with
respect to particle permutations. The above states are a (non-orthonormal) basis of the Hilbert
space of $H$ provided that (for instance) $\ell_i\ge \ell_{i+1}$ for all $i$ and $s_i\ge s_{i+1}$
whenever $\ell_i=\ell_{i+1}$. It was shown in Ref.~\cite{FGGRZ01} that $H_0$ acts triangularly on the
latter basis, with eigenvalues
\begin{equation}
  \Eg+2a\hspace{.7pt}|\bll|\,,\qquad|\bll|\equiv\sum_{i=1}^N\ell_i\,.
  \label{eigH0}
\end{equation}
On the other hand, from the identities
\[
S_{ij}S_{i}^\al=S_{j}^\al S_{ij}\,,\qquad S_{ij}S_{j}^\al=S_{i}^\al S_{ij}\,;
\qquad S_{ij}S_{k}^\al=S_{k}^\al S_{ij}\,,\quad k\ne i,j\,,
\]
it immediately follows that the operators $S^\al$ commute with the spin permutation operators
$S_{ij}$, and hence with $\La$. By Eq.~\eqref{Skal} we then have
\[
S^\al\psi_{\bll,\bbs}(\bx)=\rho(\bx)\La\Big(\,{\textstyle\prod\limits_{i=1}^N}
x_i^{\ell_i}\,S^\al\ket{s_1,\dots,s_N}\Bigr)=c^\al(\bbs)\psi_{\bll,\bbs}\,,
\]
where the eigenvalue $c^\al(\bbs)$ is given by
\[
c^\al(\bbs)=\sum_{i=1}^N\bigl(\de_{s_i}^\al-\de_{s_i}^m\bigr)\equiv
\#\{s_i=\al\}-\#\{s_i=m\}\,.
\]
In the previous formula, the symbol $\#\{s_i=k\}$ denotes the number of components of the vector
$\bbs$ equal to $k$. It follows that the term $\sum_{\al=1}^{m-1}B_\al S^\al$ in Eq.~\eqref{H} is
diagonal in the basis~\eqref{psins}, with eigenvalues
\begin{equation}
  \label{eigS}
 \sum_{\al=1}^{m-1}c^\al(\bbs)
 B_\al=\sum_{\al=1}^{m-1}B_\al\,\#\{s_i=\al\}-\#\{s_i=m\}\sum_{\al=1}^{m-1}B_\al\,.
\end{equation}
In view of the latter equation, it is convenient to introduce the notation
\begin{equation}
  \label{Bm}
  B_m=-\sum_{\al=1}^{m-1}B_\al\,,
\end{equation}
so that $\sum_{\al=1}^mB_\al=0$. Using this notation we can rewrite the eigenvalue~\eqref{eigS} in
the more compact form
\begin{equation}\label{eigSshort}
\sum_{\al=1}^mB_\al\,\#\{s_i=\al\}\equiv\sum_{i=1}^NB_{s_i}\,.
\end{equation}
From the previous considerations and Eqs.~\eqref{eigH0}-\eqref{eigSshort} it then follows that
the Hamiltonian~\eqref{H} acts triangularly on the basis~\eqref{psins}, with eigenvalues
\begin{equation}
  \label{eigH}
  E_{\bll,\bbs}=\Eg+\frac{2a}J\bigg(J\hspace{.7pt}|\bll|-\sum_{i=1}^N B_{s_i}\bigg)\,.
\end{equation}

Using the previous equation, it is a straightforward matter to compute the partition function of
the spin Calogero model~\eqref{H}. To this end, let us represent the multiindex $\bll$ in
Eq.~\eqref{bl} as
\begin{equation}
  \bll=\big(\overbrace{\la_1,\dots,\la_1}^{k_1},\dots,\overbrace{\la_r,\dots,\la_r}^{k_r}\big)\,,
  \label{bl}
\end{equation}
with $\la_1>\la_2>\cdots>\la_r\ge0$ and $(k_1,\dots,k_r)\equiv\bk\in\cP_N$. We then have
\begin{equation}
  Z(2aT/J)=q^{\frac{J\Eg}{2a}}\sum_{\bk\in\cP_N}\sum_{\la_1>\cdots>\la_r\ge0}
  q^{\sum_{i=1}^rJk_i\la_i}\sum_{\bbs\in\bll}q^{-\sum_{i=1}^NB_{s_i}}\,,
\label{Z2aTJ}
\end{equation}
where the last sum is extended to all spin quantum numbers $\bbs$ compatible with the multiindex
$\bll$ given by Eq.~\eqref{bl}, i.e., such that $s_i\ge s_{i+1}$ whenever $l_i=l_{i+1}$. In fact,
since the latter sum clearly depends on $\bll$ only through $\bk$, from now on we denote it by
$\Si(\bk)$. Clearly, by Eq.~\eqref{bl} we have
\begin{equation}
  \label{Sik}
  \Si(\bk)
  =\prod_{i=1}^r\sum_{s_1\ge\cdots\ge s_{k_i}}q^{-\sum_{j=1}^{k_i}B_{s_j}}
  \,.
\end{equation}
With this notation, Eq.~\eqref{Z2aTJ} becomes
\[
Z(2aT/J)=q^{\frac{J\Eg}{2a}}\sum_{\bk\in\cP_N}\Si(\bk)\sum_{\la_1>\cdots>\la_r\ge0}
q^{\sum_{i=1}^rJ k_i\la_i}\,.
\]
The inner sum in the RHS is easily computed by performing the change of indices
$\nu_i=\la_i-\la_{i+1}$, where $i=1,\dots,r$ and $\la_{r+1}\equiv0$, since the new indices
independently range from $1-\de_{ir}$ to $\infty$. We thus obtain
\begin{equation}
  \label{inners}
  \sum_{\la_1>\cdots>\la_r\ge0}q^{\sum_{i=1}^rJ k_i\la_i}=(1-q^{NJ})^{-1}
  \prod_{i=1}^{r-1}\frac{q^{J K_i}}{1-q^{J K_i}}\,,
\end{equation}
and therefore
\begin{equation}
  \label{Zfinal}
  Z(2aT/J)=q^{\frac{J\Eg}{2a}}\sum_{\bk\in\cP_N}\Si(\bk)\,q^{\sum_{i=1}^{r-1}J
    K_i}\prod_{i=1}^r(1-q^{J K_i})^{-1}\,,
\end{equation}
where the positive integers $K_i$ were defined in Eq.~\eqref{KKp}. From the freezing trick
relation~\eqref{Zs} and Eqs.~\eqref{ZscPF}-\eqref{Zfinal}, we immediately obtain the following
explicit expression for the partition function of the PF chain~\eqref{cH}-\eqref{PFdef} in the
presence of an arbitrary $\mathrm{su}(m)$ magnetic field:
\begin{equation}
  \label{cZfinal}
  \cZ(q)=\sum_{\bk\in\cP_N}\Si(\bk)\,q^{\sum_{i=1}^{r-1}J K_i}\prod_{i=1}^{N-r}(1-q^{J K'_i})\,.
\end{equation}

The previous argument can be applied with minor modifications to
both the HS and FI chains, thereby obtaining the following general expression for
the partition function of the HS-type chains~\eqref{cH}-\eqref{Jijs}:
\begin{equation}
  \label{cZall}
  \cZ(q)=\sum_{\bk\in\cP_N}\Si(\bk)\,q^{\sum_{i=1}^{r-1}J\cE(K_i)}\prod_{i=1}^{N-r}(1-q^{J\cE(K'_i)})\,.
\end{equation}
Note that when the magnetic field vanishes, by Eq.~\eqref{Sik} $\Si(\bk)$ becomes
$\prod_{i=1}^r{\textstyle\binom{m+k_i-1}{k_i}}$, and the previous expression for $\cZ$
reduces to its zero field version~\eqref{cZ0}.

A few remarks on the function $\Si(\bk)$ are now in order. Let us first note that
the latter function can be expressed as
\begin{equation}\label{Sisik}
\Si(\bk)=\prod_{i=1}^r \si(k_i)\,,
\end{equation}
where
\begin{equation}\label{sik}
\si(k)=\sum_{s_1\ge\cdots\ge s_k}q^{-\sum_{j=1}^{k}B_{s_j}}\,.
\end{equation}
The function $\si(k)$ can be considerably simplified by noting that the summation indices
$s_1,\dots,s_k$ can be expressed as
\begin{equation}
  \overbrace{\vphantom{1'}m,\dots,m}^{p_m}\,,\overbrace{\vphantom{1'}m-1,\dots,m-1}^{p_{m-1}}\,,
  \dots,\overbrace{\vphantom{1'}1,\dots,1}^{p_1}\,,
  \label{s1sk}
\end{equation}
with $\sum_{\al=1}^{m}p_\al=k$. Thus
\[
\sum_{j=1}^kB_{s_j}=\sum_{\al=1}^{m}p_\al B_\al\,,
\]
and therefore
\begin{equation}
  \label{sikred}
  \si(k)=\sum_{p_1+\cdots+p_m=k}\,\prod_{\al=1}^{m}q^{-p_\al B_\al}\,.
\end{equation}
The latter sum can be easily evaluated by noting that
\begin{equation}
  \si(k)=h_k(q^{-B_1},\dots,q^{-B_m})\,,
  \label{sikhk}
\end{equation}
where $h_k(\bx)$ denotes the complete homogeneous symmetric polynomial of degree $k$ in
$m$ variables $(x_1,\dots,x_m)\equiv\bx$, given by
\begin{equation}
  \label{ESP}
  h_k(\bx)=\sum_{p_1+\cdots+p_m=k}x_1^{p_1}\cdots x_m^{p_m}\,.
\end{equation}
Recalling that $h_k$ is the Schur polynomial associated with the single-row partition $(k)$ and
using Jacobi's definition of the latter polynomials in terms of determinants~\cite{Ma95} we
conclude that
\begin{equation}
  \label{sikfinal}
  \si(k)=\prod_{1\le i<j\le m}(q^{-B_i}-q^{-B_j})^{-1}\cdot\left|\begin{array}{ccc}
      q^{-(k+m-1)B_1}&\cdots&q^{-(k+m-1)B_m}\\
      q^{-(m-2)B_1}&\cdots&q^{-(m-2)B_m}\\
      q^{-(m-3)B_1}&\cdots&q^{-(m-3)B_m}\\
      \cdots&\cdots&\cdots\\
      q^{-B_1}&\cdots&q^{-B_m}\\
      1&\cdots&1
  \end{array}\right|\,.
\end{equation}
Expanding the determinant along the first row we obtain the alternative expression
\begin{equation}
  \label{sikfinal2}
  \si(k)=\sum_{i=1}^mq^{-(k+m-1)B_i}\prod_{{j=1\atop j\ne i}}^m(q^{-B_i}-q^{-B_j})^{-1}\,.
\end{equation}

In particular, for spin \half{} ($m=2$) we have $B_1=B=-B_2$,
and either expression for $\si(k)$ easily yields
\begin{equation}
  \label{siksu2}
  \si(k)
  =\frac{q^{(k+1)B}-q^{-(k+1)B}}{q^B-q^{-B}}=
  \frac{\sinh\mathopen{}\bigl((k+1)\be B\bigr)}{\sinh(\be B)}\,.
\end{equation}
The RHS of the latter formula can be conveniently expressed using the $q$-number notation. Recall,
to this end, that given two real numbers $x$ and $w>0$ the symmetric $w$-number $[[x]]_w$ is
define by
\[
[[x]]_w=\frac{w^{\frac{x}2}-w^{-\frac{x}2}}{w^{\frac12}-w^{-\frac12}}\,,
\]
which reduces to the ordinary number $x$ for $w=1$. We then have (in the spin~\half{} case)
\[
\si(k)=[[k+1]]_{q^{2B}}\,,
\]
so that the partition function of the spin~\half{} chains~\eqref{cH}-\eqref{Jijs} can be concisely written
as
\begin{equation}
  \label{cZsu2}
  \cZ(q)=\sum_{\bk\in\cP_N}\prod_{i=1}^r[[k_i+1]]_{q^{2B}}\cdot\kern 2pt
  q^{\sum_{i=1}^{r-1}J\cE(K_i)}\prod_{i=1}^{N-r}(1-q^{J\cE(K'_i)})\,.
\end{equation}
Note, again, that in the absence of a magnetic field the previous expression obviously reduces to
Eq.~\eqref{cZ0} with $m=2$.

\section{The associated vertex models}\label{sec.vms}

In this section we shall prove that the spectrum of the HS chains~\eqref{cH}-\eqref{Jijs} coincides with
that of a classical inhomogeneous vertex model, which differs from the one in Eq.~\eqref{energies}
by the addition of a magnetic field term. Our approach is based on the notion of generating
function for the partition function, as used in Ref.~\cite{BBH10} for the zero magnetic field
case.

To this end, following the latter reference we define the generating function $\cF$ of the
zero-field partition function~\eqref{cZ0} as
\begin{equation}
  \label{cF}
  \cF(\bx) = \sum_{\bk\in\cP_N}\prod_{i=1}^rh_{k_i}(\bx)\cdot
  q^{\sum_{i=1}^{r-1}J\cE(K_i)}
  \prod_{i=1}^{N-r} (1-q^{J\cE(K'_i)})\,.
\end{equation}
Since
\[
h_k(1,\dots,1)={\textstyle\binom{m+k-1}{k}}\,,
\]
by Eq.~\eqref{cZ0} we have $\cZ_0(q)=\cF(1,\dots,1)$. More generally, substituting the
expression~\eqref{sikhk} for $\si(k)$ into Eqs.~\eqref{cZfinal}-\eqref{Sisik} and using the
definition of $\cF(\bx)$ we readily obtain the identity
\begin{equation}
  \label{cZcF}
  \cZ(q)=\cF(q^{-B_1},\dots,q^{-B_m})\,.
\end{equation}

Similarly, the generating function $\cF^{\mathrm V}$ for the partition function of the classical
vertex model~\eqref{energies} was defined in Ref.~\cite{BBH10} by
\begin{equation}
\cF^{\mathrm V}(\bx) = \sum_{n_1,\dots,n_N=0}^m x_1^{w_1(\bn)}\cdots x_m^{w_m(\bn)}
q^{E_0(\bn)}\,,
\label{cFV}
\end{equation}
where the nonnegative integers $w_\al(\bn)$ are given by
\[
w_\al(\bn)=\#\{n_k=\al\}\,.
\]
Again, it is obvious that the partition function $\cZ^{\mathrm V}_0$ of the
model~\eqref{energies}-\eqref{dejk} is the value of its generating function at the point
$(1,\dots,1)$. One of the fundamental results in Ref.~\cite{BBH10} is the fact that the generating
functions~\eqref{cF} and \eqref{cFV} are \emph{identically} equal, i.e.,
\begin{equation}
  \label{cFcFV}
  \cF(\bx)=\cF^{\mathrm V}(\bx)\,,\qquad\forall\bx\in\RR^m\,.
\end{equation}
Evaluating the previous identity at the point $(1,\dots,1)$ we obtain the equality of the
zero-field partition functions~$\cZ_0$ and $\cZ_0^{\mathrm V}$, which is in fact the main result
in Ref.~\cite{BBH10}. On the other hand, if we evaluate the identity~\eqref{cFcFV} at the point
$(q^{-B_1},\dots,q^{-B_m})$ and use Eq.~\eqref{cZcF} we immediately obtain
\begin{equation}
  \label{cZcZV}
  \cZ(q)=\cF^{\mathrm V}(q^{-B_1},\dots,q^{-B_m})=\sum_{n_1,\dots,n_N=0}^m
  q^{-\sum_{\al=1}^mB_\al w_\al(\bn)}q^{E_0(\bn)}\,.
\end{equation}
Taking into account that
\[
\sum_{\al=1}^mB_\al w_\al(\bn)=\sum_{i=1}^NB_{n_i}\,,
\]
we can rewrite Eq.~\eqref{cZcZV} as
\begin{equation}
  \label{cZcZVfinal}
  \cZ(q)=\sum_{n_1,\dots,n_N=0}^mq^{E_0(\bn)-\sum_{i=1}^NB_{n_i}}\equiv\cZ^{\mathrm V}(q)\,,
\end{equation}
where $\cZ^{\mathrm V}$ denotes the partition function of the inhomogeneous classical vertex model
with energies
\begin{equation}
  \label{Ebn}
  E(\bn)=E_0(\bn)-\sum_{i=1}^NB_{n_i}\,,\qquad n_k\in\{1,\dots,m\}\,,
\end{equation}
with $E_0(\bn)$ given by Eqs.~\eqref{energies}-\eqref{dejk}. Therefore, as stated at the
beginning of this section, the spectrum of the spin chains~\eqref{cH}-\eqref{Jijs} is identical to that
of the classical vertex models defined by Eq.~\eqref{energies}-\eqref{Ebn}.

In the spin $1/2$ case, the previous equation can be simplified by noting that
$\de(j,k)$ in Eq.~\eqref{dejk} can be expressed as
\begin{equation}\label{dejk12}
\de(j,k)=(j-1)(2-k)\,,\qquad j,k=1,2,
\end{equation}
and similarly
\begin{equation}
B_j=(3-2j)B\,,\qquad j=1,2.
\end{equation}
Introducing the spin variables $\si_k=3-2n_k$, Eq.~\eqref{Ebn} becomes
\begin{equation}\label{Ebsi0}
  E(\bsi)=\frac J4\sum_{i=1}^{N-1}\cE(i)(1-\si_i)(1+\si_{i+1})-B\sum_{1=1}^N\si_{i}\,,
  \qquad \si_k\in\{\pm1\}\,.
\end{equation}
The latter equation can be alternatively written as
\begin{equation}\label{Ebsi}
E(\bsi)=-\frac J4\sum_{i=1}^{N-1}\cE(i)(\si_i\si_{i+1}-1)-\sum_{i=1}^N\cB(i)\,\si_i\,,
\end{equation}
where
\begin{equation}
  \label{bidef}
  \cB(i)\equiv B+\frac{J}4\,\big[\cE(i)-\cE(i-1)\big]
\end{equation}
and we have set $\cE(0)=\cE(N)=0$. The last two equations define a classical {\em inhomogeneous}
Ising model in one dimension, where the coupling between the spins $i$ and $i+1$ is proportional
to the dispersion relation $\cE(i)$, and the external magnetic field (also inhomogeneous) is given
by Eq.~\eqref{bidef}.

\section{Thermodynamics of the spin \half{} chains}\label{sec.thermo}

We shall next take advantage of the representation~\eqref{Ebsi0} of the spectrum of the HS-type
chains~\eqref{cH}-\eqref{Jijs} with spin $\half{}$ in order to determine their equilibrium thermodynamics
in a unified way. To this end, we must first normalize the Hamiltonian~\eqref{cH2} to ensure that
the energy per spin is finite (and nonzero) in the thermodynamic limit. It is well known that when
$N\gg1$ the mean energy of the HS-type chains is $\Or(N^{r})$, with $r=2$ for the PF chain and
$r=3$ for the HS and FI chains~\cite{FG05,BFGR08epl,BFGR10}. Hence from now on we shall take
\begin{equation}
  \label{Kdef}
  J=\begin{cases}
    K/N^2\,, \quad &\text{for the HS and FI chains}\\
    K/N\,,& \PF,
  \end{cases}
\end{equation}
where the constant $K$ is independent of $N$. With this proviso, Eq.~\eqref{Ebsi0} can be rewritten
in terms of the discrete variables $x_i\equiv i/N\in(0,1)$ as
\begin{equation}
  \label{EbsiTL}
  E(\bsi)=\frac K4\sum_{i=1}^{N-1}\vep_i(1-\si_i)(1+\si_{i+1})-B\sum_{i =1}^N\si_{i}\,,
\end{equation}
where
\begin{equation}\label{epi}
\vep_i=\begin{cases}
  x_i(1-x_i)\,,&\HS\\
  x_i\,,&\PF\\
  x_i(\ga_N+x_i)\,,\quad &\FI,
\end{cases}
\end{equation}
and $\ga_N\equiv(\al-1)/N$. We shall further assume that $\ga_N$ has a finite limit
$\ga\ge0$ as $N\to\infty$. With the above notation, the partition function of the
chains~\eqref{cH2}-\eqref{Jijs} can be collectively expressed as
\begin{equation}\label{Zlong}  
\cZ(q)=\sum_{\bsi}q^{d(\si_1,\si_2)\vep_1-\frac B2(\si_1+\si_2)}\,\cdots\,
q^{d(\si_{N-1},\si_N)\vep_{N-1}-\frac B2(\si_{N-1}+\si_N)}q^{-\frac B2(\si_1+\si_N)}\,,
\end{equation}
where
\[
d(\si,\si')=\frac K4(1-\si)(1+\si')\,.
\]
Equation~\eqref{Zlong} can be more concisely written as
\begin{equation}
  \label{Ztr}
  \cZ(q)=\tr(UT_1\cdots T_{N-1})\,,
\end{equation}
where $U$ and $T_i$ are $2\times2$ matrices with elements
\begin{equation}\label{UT}
  U_{\si\si'}=q^{-\frac B2(\si+\si')}\,,\quad (T_i)_{\si\si'}=q^{d(\si,\si')\vep_i-\frac B2(\si+\si')}\,;
  \qquad \si,\si'=\pm1\,,
\end{equation}
or equivalently,
\begin{equation}
  \label{Utmatrix}
  U=
  \left(\begin{array}{cc}
    q^B& 1\\1& q^{-B}
  \end{array}\right)\,,\qquad
T_i=\left(\begin{array}{cc}
    q^B& q^{K\vep_i}\\1& q^{-B}
  \end{array}\right)\,.
\end{equation}
The transfer matrix $T_i$ has two distinct eigenvalues
\begin{equation}
  \la_{i,\pm}=\cosh(\be B)\pm\sqrt{\sinh^2(\be B)+\e^{-K\be\vep_i}}\,,
  \label{laipm}
\end{equation}
and is therefore diagonalizable. A straightforward calculation shows that
\begin{equation}
  T_i=P_iD_iP_i^{-1}\,,
\end{equation}
where
\begin{equation}
D_i=\left(
  \begin{array}{cc}
    \la_{i,+}& 0\\0& \la_{i,-}
  \end{array}
\right)\,,\qquad P_i=\left(
  \begin{array}{cc}
    -a+r_i& \enspace -a-r_i\\[2mm]1& \enspace 1
  \end{array}
\right)\,,
\label{DiPi}
\end{equation}
and we have set
\begin{equation}
  a=\sinh(\be B)\,,\qquad r_i=\sqrt{\sinh(\be B)^2+\e^{-K\be\vep_i}}\,.
  \label{abi}
\end{equation}
We thus have
\[
T_1\cdots T_{N-1}=P_1(D_1C_1)\cdots(D_{N-2}C_{N-2})D_{N-1}P_{N-1}^{-1}\,,
\]
where $C_i\equiv P_i^{-1}P_{i+1}$ is a symmetric matrix explicitly given by
\begin{equation}\label{Ci}
  C_i=\frac1{2r_i}\,\left(
    \begin{array}{cc}
      r_i+r_{i+1}& r_i-r_{i+1}\\[2mm]r_i-r_{i+1}& r_i+r_{i+1}
    \end{array}
  \right)\,.
\end{equation}
The key observation at this point is that
\begin{equation}
  C_i=I+\Or(N^{-1})
  \label{CiO}
\end{equation}
as $N\to\infty$. Indeed, note first of all that
\[
0\le\vep_i\le\vep_{\rm max}\,,
\]
where
\begin{equation}
\vep_{\rm max}=\begin{cases}\frac14\,,&\HS\\
    1\,,&\PF\\\ga_N+1\,,&\FI,
  \end{cases}
\end{equation}
so that $\vep_i=\Or(1)$. On the other hand, since $0<x_i<1$ the difference
\[
\vep_{i+1}-\vep_i=\begin{cases}
  \frac1N\,(1-x_i-x_{i+1})\,,&\HS\\
  \frac1N\,,&\PF\\
  \frac1N\,(\ga_N+x_i+x_{i+1})\,,&\FI
\end{cases}
\]
is $\Or(N^{-1})$, and hence
\[
\frac{r_{i+1}}{r_i}=1+\Or(N^{-1})\,,
\]
which establishes our claim.

{}From Eq.~\eqref{CiO}, it follows that when $N\to\infty$ the leading term of $\cZ(q)$ is given by
\[
\tr\mathopen{}\left(P_{N-1}^{-1}UP_1D_1\cdots D_{N-1}\right)=\tr\left[P_{N-1}^{-1}UP_1\left(
    \begin{array}{cc}
      \la_+&0\\0&\la_-
    \end{array}\right)\right]\,,
\]
where
\[
\la_\pm\equiv\prod_{i=1}^{N-1}\la_{i,\pm}\,.
\]
Note also that
\[
\lim_{N\to\infty}\vep_1=0\,,\qquad \lim_{N\to\infty}\vep_{N-1}=\vep_{\infty}\equiv\begin{cases}
  0\,,&\HS\\1\,,&\PF\\\ga+1\,,\quad&\FI,
\end{cases}
\]
so that the matrices $P_1$ and $P_{N-1}$ both have finite limits as $N\to\infty$. Calling
\[
r_\infty=\sqrt{\sinh^2(\be B)+\e^{-K\be\vep_\infty}}
\]
we then have
\begin{align}
  \lim_{N\to\infty} P_{N-1}^{-1}UP_1&=\frac1{2r_\infty}\left(
    \begin{array}{cc}
      1&r_\infty+a\\-1&r_\infty-a
    \end{array}
  \right)
  \left(
    \begin{array}{cc}
      \e^{-\be B}&1\\
      1& \e^{\be B}
    \end{array}
  \right) \left(
    \begin{array}{cc}
      \e^{-\be B}& -\e^{\be B}\\
      1& 1\\
    \end{array}
  \right)\nonumber\\
  \label{MP1}
  &=\frac{\cosh(\be B)}{r_\infty}\,\left(
    \begin{array}{cc}
     r_\infty+\cosh(\be B) & 0\\r_\infty-\cosh(\be B)& 0
    \end{array}
  \right)\,,
\end{align}
and therefore
\[
\tr\left[P_{N-1}^{-1}UP_1\left(
    \begin{array}{cc}
      \la_+&0\\0&\la_-
    \end{array}\right)\right]=\cosh(\be B)\left(1+\frac{\cosh(\be B)}{r_\infty}\right)\la_+
\]
is the leading term of $\cZ(q)$ as $N\to\infty$. Since the first two factors in the RHS of this
formula are independent of $N$, in the thermodynamic limit the free energy per spin $f(B,T)$ is
simply given by
\begin{align}
  \hspace{1.2cm}f(B,T)&=
  -\lim_{N\to\infty}\frac{\log\cZ}{N\be}=-\frac1{\be}\lim_{N\to\infty}N^{-1}
  \sum_{i=1}^{N-1}\log\la_{i,+}\nonumber\\&=-\frac1{\be}\lim_{N\to\infty}N^{-1}
  \sum_{i=1}^{N-1}\log\Bigl(\cosh(\be B)+\sqrt{\sinh^2(\be B)+\e^{-K\be\vep(x_i)}}\,\Bigr)\,,
   \label{flim}
\end{align}
where the \emph{dispersion function} $\vep(x)$ is defined by
\begin{equation}\label{disprel}
\vep(x)=\begin{cases}
  x(1-x)\,,& \HS\\
  x\,,&\PF\\
  x(\ga+x)\,,\qquad&\FI
\end{cases}
\end{equation}
(cf.~Eq.~\eqref{epi}). We thus obtain the following remarkable formula for the free energy per site
of the HS-type chains~\eqref{cH2}-\eqref{Jijs}:
\begin{equation}\label{f}
  f(B,T)=-\frac1\be\int_0^1\log\Bigl(\cosh(\be B)+\sqrt{\sinh^2(\be B)+\e^{-K\be\vep(x)}}\,\Bigr)\,\diff x.
\end{equation}
From the latter equation it is immediate to obtain the magnetization per site:
\begin{equation}\label{M}
  \mu(B,T)=-\frac{\partial f(B,T)}{\partial B}=\int_0^1 \frac{\sinh(\be B)}{\sqrt{\sinh^2(\be B)+\e^{-K\be\vep(x)}}}\,\diff x\,.
\end{equation}
For finite temperature, the magnetic susceptibility is also easily computed from the previous
formula:
\begin{equation}\label{chi}
  \chi(B,T)=\frac{\partial\mu(B,T)}{\partial B}=\be\int_0^1\frac{\cosh(\be
    B)\e^{-K\be\vep(x)}}{\left(\sinh^2(\be B)+\e^{-K\be\vep(x)}\right)^{3/2}}\,\diff x\,.
\end{equation}
In the case of the FI chain, some of the above equations have previously appeared in
Ref.~\cite{FI94}. More precisely, Eq.~\eqref{f} with $B=0$ reduces to Eq.~(21) of the latter
reference, while Eq.~\eqref{M} for $K>0$ essentially coincides with Eq.~(24) of Ref.~\cite{FI94},
which is stated without proof. As first noted in this reference for the $B=0$ case, Eqs.~\eqref{f}--\eqref{chi} imply that the HS chain is thermodynamically equivalent to an FI chain with $\ga=-1$
and $K\to-K$.

The free energy and magnetization per spin~\eqref{f}-\eqref{M} admit a simpler
expression in terms of the so-called
{\em dressed energy} $\vep_{\mathrm{dr}}(x)$ defined by
\begin{equation}\label{vepdr}
\sinh(\be\vep_{\mathrm{dr}}(x))=\e^{\frac{K\be}2\vep(x)}\sinh(\be B)\,,
\end{equation}
namely,
\begin{align}
  f(B,T)&=
  -\frac1\be\int_0^1\log\mathopen{}\left(\frac{\sinh[\be(\vep_{\mathrm{dr}}(x)+B)]}{\sinh[\be\vep_{\mathrm{dr}}(x)]}\right)\diff x\label{fvepdr}\\[1mm]
  \mu(B,T)&=\int_0^1\tanh(\be\vep_{\mathrm{dr}}(x))\,\diff x\,.
\end{align}  
These are the well-known expressions for the free energy (up to an additive constant depending on
the normalization of $\cH$) and magnetization per particle of a system of {\em free} bosons with
one-particle energies $\vep(x)$, $x\in[0,1]$, in the presence of a constant magnetic field $B$.
Note in this respect that, in contrast with the dispersion relation $\vep(x)$, the dressed energy
$\vep_{\mathrm{dr}}(x)$ depends on $\be$, $B$ and $K$. Note also that in the case of
the HS chain Eq.~\eqref{fvepdr} is essentially equivalent to Eq.~(37) in Ref.~\cite{BPS93},
deduced by means of Haldane's spinon gas approach~\cite{Ha91}.

For future reference, we shall also compute the internal energy and entropy per spin. To begin
with, the internal energy per spin $u$ is given by
\begin{equation}\label{u}
   u(B,T)=\frac{\partial}{\partial\be}\Big(\be f(B,T)\Big)=
  \frac12\int_0^1\left(\frac{K\vep(x)\,\e^{-K\be\vep(x)}}{\cosh(\be B)+r(x)}-2B\sinh(\be
    B)\right)\frac{\diff x}{r(x)}\,,
\end{equation}
with
\[
r(x)=\sqrt{\sinh^2(\be B)+\e^{-K\be\vep(x)}}\,.
\]
The entropy per spin $s$ is then easily derived from the usual formula
\begin{equation}\label{entropy}
s(B,T)=\frac{u(B,T)-f(B,T)}T\,.
\end{equation}

\subsection{The PF chain}
\label{ssec.PF}

We have not been able to express the thermodynamic functions~\eqref{f}--\eqref{entropy} in terms of
known special functions for arbitrary values of $B$ and $T$, with the remarkable exception of the
Polychronakos--Frahm chain. Indeed, we shall now show that in this case the free and the internal
energy, the entropy (and the specific heat) can be expressed in terms of the dilogarithm
function~\cite{Le81}
\begin{equation}\label{Li2}
\Li_2(z)=-\int_0^z\frac{\log(1-t)}{t}\,\diff t\,,\qquad z\in\CC\setminus(1,\infty)\,,
\end{equation}
while the magnetization and susceptibility are elementary functions of $B$ and $T$.
In Eq.~\eqref{Li2} $\log$ denotes the principal value of the logarithm (i.e., $-\pi<\im\log z<\pi$),
and the integral is taken along any path joining the origin with the point $z$ that does not
intersect the branch cut $(1,\infty)$.

We start by performing the change of variable $t^2=a^2+\e^{-\ka x}$, where
$a=\sinh(\be B)$ and $\ka=K\be$,
in Eq.~\eqref{f}
for the free energy. We readily obtain
\[
K\be^2 f(B,T)=\int_c^{\sqrt{a^2+\e^{-\ka}}}\left(\frac1{t+a}+\frac1{t-a}\right)\log(t+c)\,\diff
t\,,
\]
where $c=\cosh(\be B)$. The linear change of variable
\[
z = -\frac{t\pm a}{c\mp a}\equiv-\e^{\pm\be B}(t\pm a)
\]
makes it possible to evaluate each of the last two integrals in closed form:
\begin{align*}
\int_c^{\sqrt{a^2+\e^{-\ka}}}\frac{\log(t+c)}{t\pm a}\,\diff t
&=\int_{-\e^{\pm2\be B}}^{-\e^{\pm\be B}(\pm a+\sqrt{a^2+\e^{-\ka}})}
\left(\frac{\log(1-z)}{z}\mp\frac{\be B}{z}\right)\diff z\\
&=\Li_2(-\e^{\pm2\be B})-\Li_2\Big(-\e^{\pm\be B}(\pm a+\sqrt{a^2+\e^{-\ka}})\Big)\\
&\hphantom{{}=}\mp\be B\log(\pm
a+\sqrt{a^2+\e^{-\ka}})+\be^2 B^2\,.
\end{align*}
Taking into account the dilogarithm identity~\cite{Le81}
\begin{equation}\label{dilogid}
\Li_2(-\e^x)+\Li_2(-\e^{-x})=-\frac{\pi^2}6-\frac{x^2}2\,,
\end{equation}
after some straightforward algebra we finally obtain the following closed formula for the free
energy of the PF chain:
\begin{align}
  -K\be^2\Big(f(B,T)+B\Big)&=\frac{\pi^2}6+2\be B\log\Bigl(\sinh(\be B)+\sqrt{\sinh^2(\be
    B)+\e^{-K\be}}\,\Bigr)\nonumber\\
  &\hphantom{{}=}+\Li_2\Bigl(-\e^{\be B}\Big[\sinh(\be B)+\sqrt{\sinh^2(\be
    B)+\e^{-K\be}}\,\Big]\Bigr)\nonumber\\
  &\hphantom{{}=}+\Li_2\Bigl(-\e^{-\be B}\Big[-\sinh(\be B)+\sqrt{\sinh^2(\be
    B)+\e^{-K\be}}\,\Big]\Bigr).
  \label{fPF}
\end{align}
The previous formula generalizes the analogous expression for zero magnetic field in Eq.~(23) of
Ref.~\cite{Fr93}. Equation~\eqref{fPF} can be somewhat simplified by expressing the RHS
in terms of the dressed energy. Indeed, by Eq.~\eqref{vepdr} with $\vep(x)=x$ we have
\begin{equation}\label{dressedvep}
\pm\sinh(\be B)+\sqrt{\sinh^2(\be B)+\e^{-K\be}}=\e^{\be\left(-\frac K2\pm\vep_{\rm dr}(1)\right)}\,,
\end{equation}
and therefore
\[
-K\be^2f(B,T)=\frac{\pi^2}6+2\be^2B\,\vep_{\rm dr}(1)+\Li_2\Bigl(-\e^{\be\left(B+\vep_{\rm
      dr}(1)-\frac K2\right)}\Bigr) +\Li_2\Bigl(-\e^{-\be\left(B+\vep_{\rm dr}(1)+\frac
    K2\right)}\Bigr)\,.
\]

From the usual formulas for the internal energy, the entropy and the specific heat (per site)
\[
c(B,T)=\frac{\partial u(B,T)}{\partial T}
\]
in terms of the free energy, it immediately follows that these functions necessarily include a
term proportional to the last two lines in Eq.~\eqref{fPF}, and therefore cannot be expressed
purely in terms of elementary functions. On the other hand, since the magnetization is obtained
from the free energy by differentiating with respect to the magnetic field $B$, it is clear from
Eq.~\eqref{fPF} that $\mu$ is an elementary function of $B$ and $T$. Differentiating Eq.~\eqref{fPF}
with respect to $B$, or simply performing the change of variables $t^2=a^2+\e^{-K \be x}$ in
Eq.~\eqref{M}, after a straightforward calculation we obtain the remarkable formula
\begin{equation}\label{MPF}
  \mu(B,T)=1+\frac2{K\be}\left[\log\Bigl(\sinh(\be B)+\sqrt{\sinh^2(\be
    B)+\e^{-K\be}}\,\Bigr)-\be B\right].
\end{equation}
Using Eq.~\eqref{dressedvep}, we can rewrite the previous formula in terms of the dressed energy as
\begin{equation}
  \label{MPF3}
  \mu(B,T)=\frac2K\Big(\vep_{\rm dr}(1)-B\Big)\,.
\end{equation}
Differentiating the previous expressions for $\mu$ we also obtain the following closed
formulas for the susceptibility per site:
\begin{equation}\label{chiPF}
  \chi(B,T)=\frac2K\left(\frac{\cosh(\be B)}{\sqrt{\sinh^2(\be B)+\e^{-K\be}}}-1\right)
  =\frac2K\left(\frac{\tanh\bigl(\be \vep_{\rm dr}(1)\bigr)}{\tanh(\be B)}-1\right).
\end{equation}

\section{The zero field case}\label{sec.zeroB}

It is of interest to study the limit of the thermodynamic functions computed in the previous
section for both $B\to0$ and $T\to0$. We shall deal in this section with the first of these
limits, leaving the analysis of the zero temperature limit for the next one.

We shall start with the free energy and the internal energy per site, which can be immediately
obtained by setting $B=0$ in Eqs.~\eqref{f} and~\eqref{u}:
\begin{align}
f(0,T)&=-\frac1\be\int_0^1\log\Bigl(1+\e^{-\frac{K\be}2\vep(x)}\Bigr)\diff x\,,\label{f0T}\\[1mm]
u(0,T)&=\frac K2\int_0^1\frac{\vep(x)}{1+\e^{\frac{K\be}2\vep(x)}}\,\diff x\label{u0T}\,.
\end{align}
{}From the above formulas and Eq.~\eqref{entropy}  it follows that the entropy per spin
is given by
\begin{align}
  \frac {s(0,T)}{\kB}&=\int_0^1\Bigg[\log\mathopen{}\left(1+\e^{-\frac{K\be}2\vep(x)}\right)
  +\frac12\,\frac{K\be\vep(x)}{1+\e^{\frac{K\be}2\vep(x)}}
  \Bigg]\diff x\nonumber\\[1mm]
  &= \int_0^1\left\{\log\mathopen{}\left[2\cosh\mathopen{}\left(\case{K\be}4\vep(x)\right)\right]
    -\frac{K\be\vep(x)}4\,\tanh\mathopen{}\left(\case{K\be}4\vep(x)\right)\right\}\diff x\,.
  \label{s0T}
\end{align}
In the case of the HS chain, the latter equation is essentially equivalent to Eq.~(15) in
Ref.~\cite{Ha91}, which was deduced using the equivalence of the HS chain to an ideal gas of
spinons. The zero-field specific heat
\[
c(0,T)\equiv\frac{\partial u(0,T)}{\partial T}
\]
can also be easily computed from Eq.~\eqref{u0T}, with the result
\begin{equation}\label{c0T}
\frac {c(0,T)}{\kB}=\left(\frac{K\be}4\right)^2\int_0^1
\vep^2(x)\sech^2\mathopen{}\left(\case{K\be}4\vep(x)\right)\diff x\,.
\end{equation}
Obviously, the zero-field magnetization vanishes (cf.~\eqref{M}), while Eq.~\eqref{chi} for
the susceptibility reduces to
\begin{equation}\label{chi0T}
\chi(0,T)=\be\int_0^1\e^{\frac{K\be}2\vep(x)}\diff x\,.
\end{equation}
In fact, the zero-field susceptibility can be expressed in terms of standard special functions for
all spin chains of HS type. For the PF chain, we have already obtained a formula for $\chi$ in terms
of elementary functions for all values of $B$ and $T$, cf.~Eq.~\eqref{chiPF}. For the HS chain, a
straightforward calculation shows that
\begin{equation}\label{chi0THS}
  \chi(0,T)=\sqrt{\frac{2\pi\be}{|K|}}\,\e^{K\be/8}\,\times
  \begin{cases}\erf\biggl(
    \textstyle\sqrt{\frac{K\be}8}\,\biggr)\vrule width0pt depth10pt\,,& $K>0$\\
    \erfi\biggl(
    \textstyle\sqrt{\frac{|K|\be}8}\,\biggr)\,,& $K<0$\,,
  \end{cases}
\end{equation}
where
\[
\erf(z)=\frac{2}{\sqrt\pi}\int_0^z\e^{-t^2}\,\diff t\,,\qquad
\erfi(z)=\frac{2}{\sqrt\pi}\int_0^z\e^{t^2}\diff t=-\iu\erf(\iu z)
\]
denote respectively the error and the imaginary error functions~\cite{AS70}.
Similarly, for the FI chain we have
\begin{equation}
  \label{chi0TFI}
  \chi(0,T)=\sqrt{\frac{\pi\be}{2|K|}}\,\e^{-K\be\ga^2/8}\,\times
\begin{cases}\erfi\biggl(
    (\ga+2)\textstyle\sqrt{\frac{K\be}8}\,\biggr)
    -\erfi\bigg(\ga\textstyle\sqrt{\frac{K\be}8}\,\bigg)\vrule width0pt depth12pt,& K>0\\
\erf\biggl(
    (\ga+2)\textstyle\sqrt{\frac{|K|\be}8}\,\biggr)
    -\erf\biggl(\ga\textstyle\sqrt{\frac{|K|\be}8}\,\biggr),& K<0\,.
  \end{cases}
\end{equation}
{}From Eqs.~\eqref{chiPF}, \eqref{chi0THS}, and \eqref{chi0TFI}, and the well-known
asymptotic formula~\cite{AS70}
\[
\erfi z\sim \frac{\e^{z^2}}{\sqrt\pi z}\,;\qquad |z|\to\infty\,,\quad |\arg z|<\frac\pi4\,,
\]
it follows that when $K>0$ and $T\to0$ we have
\smallskip
\[
\chi(0,T)\simeq\begin{cases}\sqrt{\frac{2\pi\be}K}\,\e^{K\be/8}\,,&\HS\vrule width0pt depth12pt\\
  \frac2K\,\e^{K\be/2}\,,&\PF\vrule width0pt depth12pt\\
  \frac2{(\ga+2)K}\,\e^{(\ga+1)K\be/2}\,,&\FI\,.
\end{cases}
\]
\smallskip Again, in the case of the HS chain the previous asymptotic formula was first obtained
by Haldane in Ref.~\cite{Ha91}. {}From the latter equation it is apparent that the ferromagnetic
zero-field susceptibility (and hence the correlation length) is exponentially divergent as
$T\to0$. This shows that all ferromagnetic chains of HS type undergo a phase transition at
$T=B=0$, which we shall analyze in detail in the next section.

Equations~\eqref{s0T} and~\eqref{c0T} imply that the zero-field entropy and specific
heat are even functions of the coupling constant $K$. This fact, which is not obvious a
priori, was first noted by Haldane~\cite{Ha91} for the entropy of the original HS chain. Haldane
also pointed out that (with an appropriate choice of the ground state energy, which amounts to
replacing $1-S_{ij}$ by $S_{ij}$ in Eq.~\eqref{cH2}) the internal energy of the HS chain is again
even in $K$. One can easily show that this property is shared by {\em all} spin chains of HS type,
since from Eq.~\eqref{u0T} it readily follows that 
\[
u(0,T;K)-u(0,T;-K)=\frac K2\int_0^1\vep(x)\,\diff x\equiv\frac{K\vep_0}2\,,
\]
with
\begin{equation}
  \label{vep0}
  \vep_0=\begin{cases}\case 1{6}\,,&\HS\vrule width0pt depth8pt\\
    \case 12\,,&\PF\vrule width0pt depth8pt\\ \case 16\,(3\ga+2)\,,&\FI.
  \end{cases}
\end{equation}
In other words, $u(0,T)-\frac {K\vep_0}4$ is an even function of $K$, which can be seen to be
equivalent to Haldane's statement given our normalization of $\cH$. A similar property, proved in
exactly the same way, also holds for the free energy $f(0,T)$.

As we saw in the previous section, in the particular case of the Polychronakos--Frahm chain the
thermodynamic functions~\eqref{f0T}--\eqref{c0T} can be expressed in closed form in terms of the
dilogarithm function. Indeed, setting $B=0$ in Eq.~\eqref{fPF} we obtain the simple formula
\begin{equation}
  \label{f0PF}
  f(0,T)=-\frac{2}{K\be^2}\bigg(\Li_2(-\e^{-K\be/2})+\frac{\pi^2}{12}\kern1pt\bigg)\,.
\end{equation}
From the latter equation it is straightforward to derive similar expressions for the internal
energy, the entropy and the specific heat of the PF chain in the absence of a magnetic field,
namely
\begin{align}
  \label{u0PF}
  u(0,T)&=\frac{2}{K\be^2}\bigg(\Li_2(-\e^{-K\be/2})+\frac{\pi^2}{12}\kern1pt\bigg)-\frac1\be\log(1+\e^{-K\be/2}\kern1pt)\\
  \frac{s(0,T)}{\kB}&=\frac{4}{K\be}\bigg(\Li_2(-\e^{-K\be/2})+\frac{\pi^2}{12}\kern1pt\bigg)-\log(1+\e^{-K\be/2}\kern1pt)\label{s0PF}\\
  \frac{c(0,T)}{\kB}&=
  \frac{4}{K\be}\bigg(\Li_2(-\e^{-K\be/2})+\frac{\pi^2}{12}\kern1pt\bigg)-2\log(1+\e^{-K\be/2}\kern1pt)-\frac{K\be}{2\,\left(1+\e^{K\be/2}\right)}\,.
  \label{c0PF}
\end{align}

\subsection{Discussion}\label{subsec.discussion}

\begin{figure}[h]
\includegraphics[height=4.3cm]{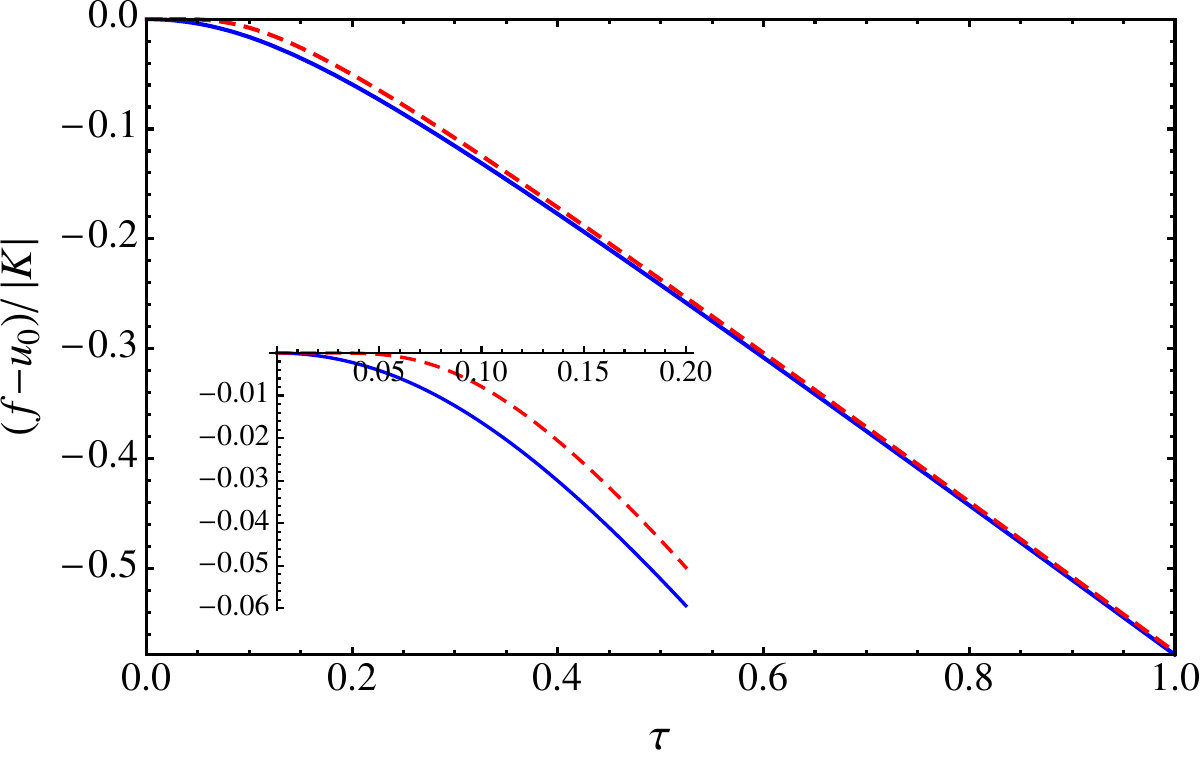}\hfill 
\includegraphics[height=4.2cm]{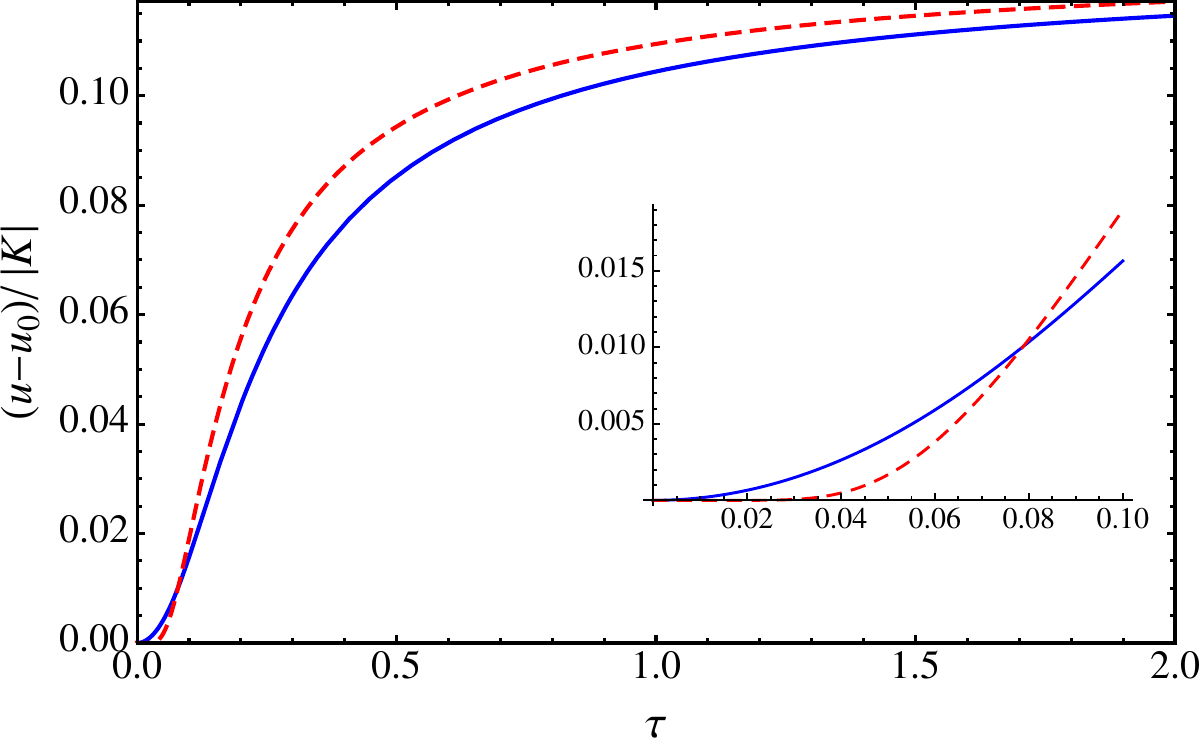}\\[1cm]
\null\kern6pt\includegraphics[height=4.2cm]{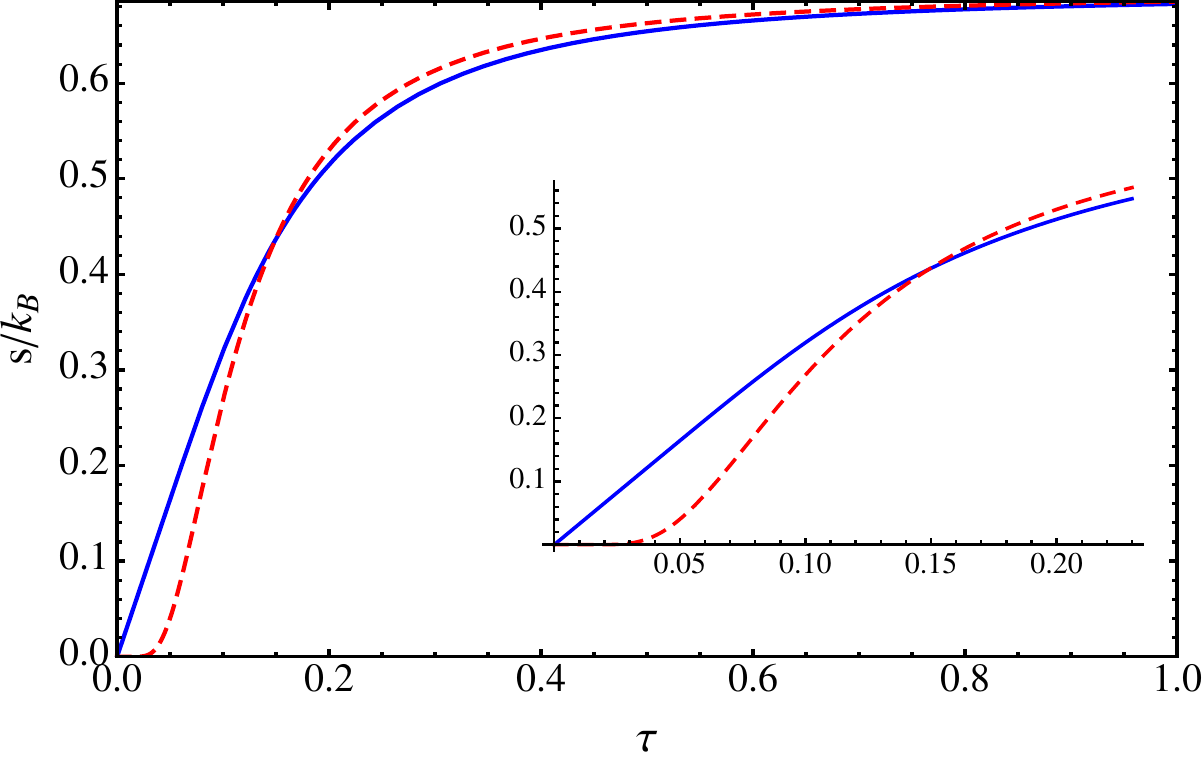}\hfill
\includegraphics[height=4.2cm]{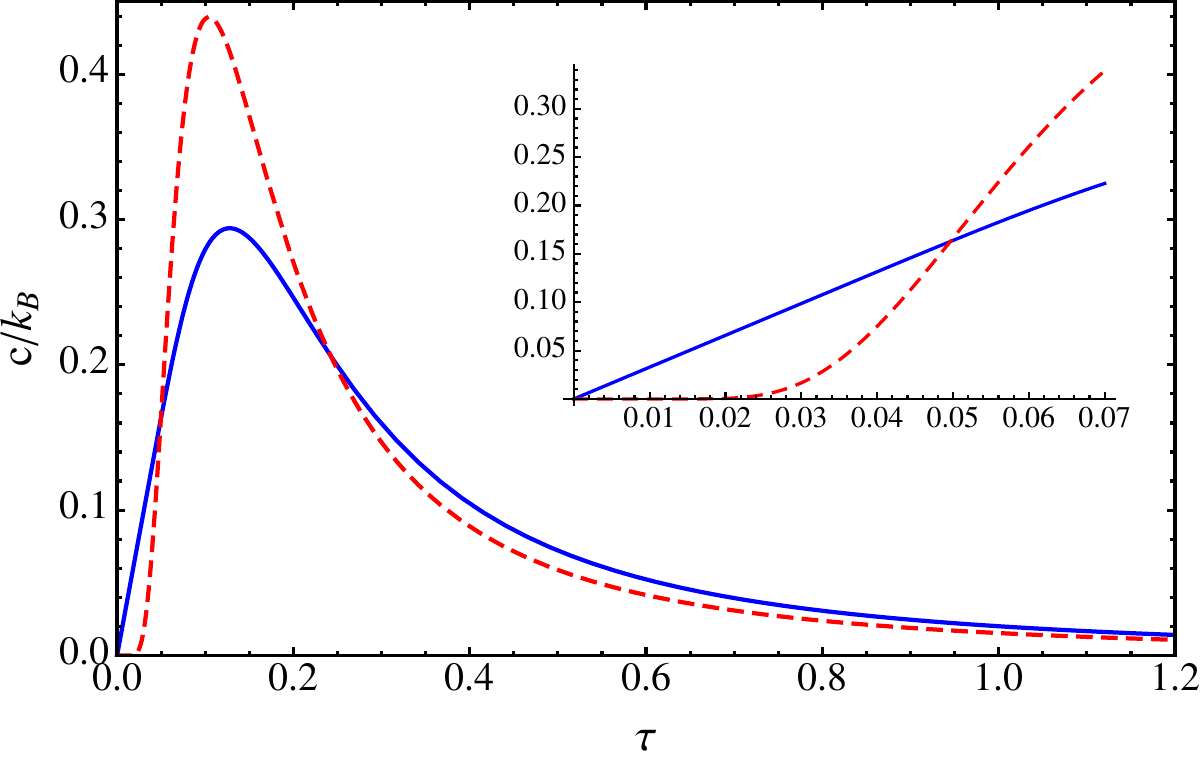}
\caption{Free and internal energy, entropy, and specific heat (all of them per site) of the PF
  chain with zero magnetic field (solid blue line) and of a two-level system with half-gap
  $\De=|K|/8$ (dashed red line) as functions of the dimensionless temperature
  $\tau=(|K|\be)^{-1}$. In the first two plots, $u_0=\frac {K\vep_0}4(1-\sgn K)$ denotes the
  common value of $u(0,0)$ and $f(0,0)$, cf.~Eq.~\eqref{u0}. Insets: low-temperature behavior of
  each of the above thermodynamic functions.}
\label{fig.PF}
\end{figure}

Using the previous explicit formulas, we have plotted the thermodynamic functions of the PF chain
for zero magnetic field~\eqref{f0PF}--\eqref{c0PF} in terms of the dimensionless temperature
$\tau=(|K|\be)^{-1}$; see Fig.~\ref{fig.PF}.

It is immediately apparent the qualitative similarity of these plots with the corresponding ones
for a two-level system like, e.g., a paramagnetic spin \half{} ion or the one-dimensional Ising
model with zero magnetic field\footnote{For the latter model, the effective half-gap is equal to
  the coupling between the spins.}~\cite{Mu10}. In particular, the specific heat $c(0,T)$ clearly
exhibits the so-called Schottky anomaly, characteristic of this type of systems. The Schottky
peak, which can be numerically computed without difficulty from Eq.~\eqref{c0PF}, is located at
$|K|\be\simeq0.127762$, with a maximum value of $c(0,T)$ approximately equal to $0.439229\,\kB$.
On the other hand, the low temperature behavior of $s(0,T)$ and $c(0,T)$ markedly differs from
that of a two-level system. Indeed, from Eqs.~\eqref{s0PF} and~\eqref{c0PF} it immediately follows
that
\begin{align}
  \frac{s(0,T)}{\kB}&=\frac{\pi^2}{3|K|\be}+\Or(T^2)\,,\qquad T\to0\,,\label{s0TPF}\\
  \frac{c(0,T)}{\kB}&=\frac{\pi^2}{3|K|\be}+\Or(T^2)\,,\qquad T\to0\,.\label{c0TPF}
\end{align}
Hence both functions have a nonvanishing first derivative at $T=0$. As is well known, the entropy
$s_\De(T)$ and the specific heat $c_\De(T)$ of a two-level system with energy gap $2\Delta>0$
satisfy
\begin{align}
\frac{s_\De(T)}{\kB}=\log(2\cosh(\be\De))-\be\De\tanh(\be\De)
\sim2\be\De\,\e^{-2\be\De}\,,\qquad T\to0\,,\label{sDe}\\
\frac{c_\De(T)}{\kB}=(\be\De)^2\sech^2(\be\De)\sim4\be^2\De^2\e^{-2\be\De}\,,\qquad T\to0\,,
\label{cDe}
\end{align}
so that the derivatives of all orders of $s_\De(T)$ and $c_\De(T)$ vanish at $T=0$.

\begin{figure}[h]
\includegraphics[height=4.25cm]{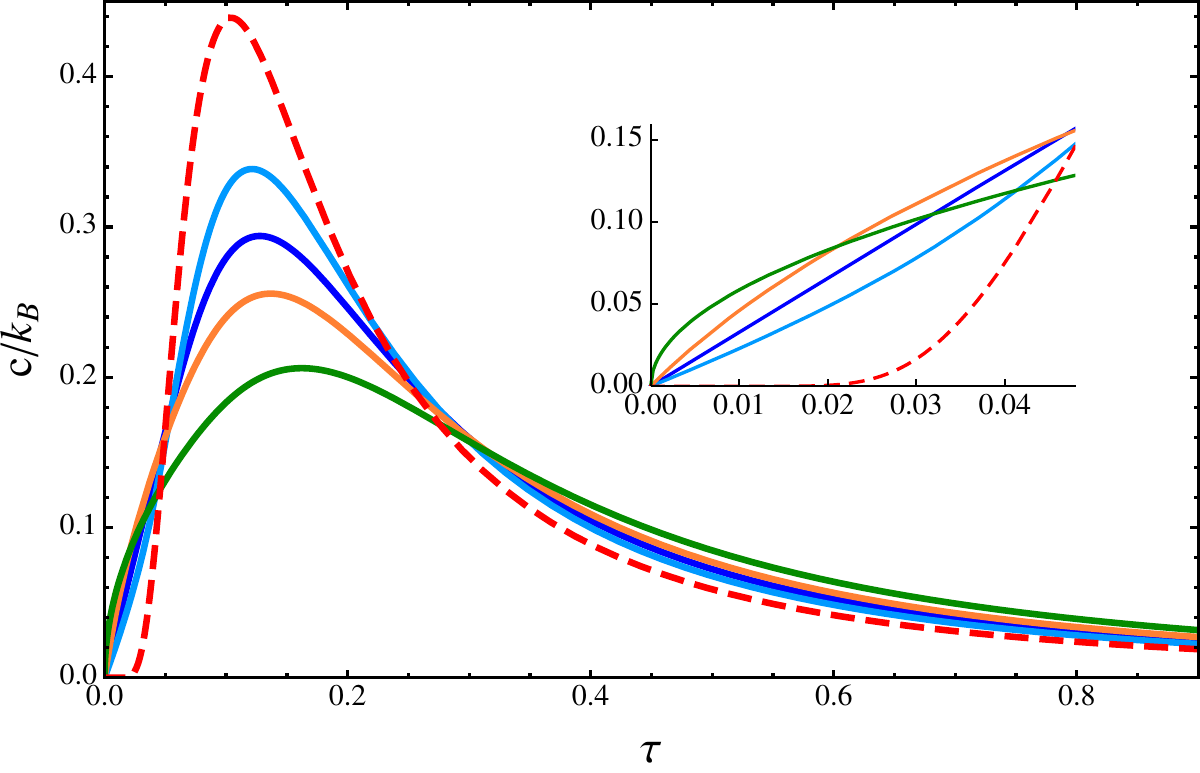}\hfill 
\includegraphics[height=4.25cm]{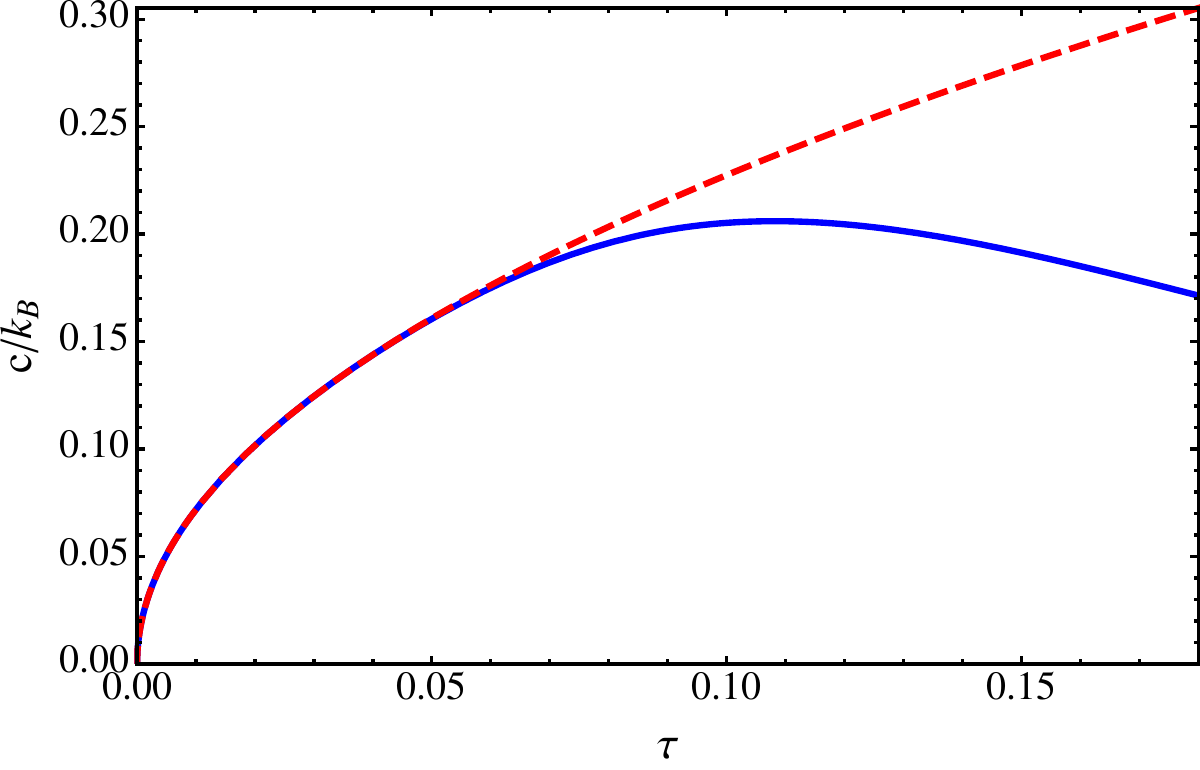}
\caption{Left: specific heat per site of the PF (blue), HS (light blue), and FI ($\gamma=1$,
  orange; $\gamma=0$, green) chains with zero magnetic field and of a two-level system with
  half-gap $\De=|K|/8$ (dashed red line) as functions of the dimensionless temperature
  $\tau=(|K|\be)^{-1}$. Right: specific heat per site of the FI
  chain with $\ga=0$ (solid blue line) versus its low-temperature approximation~\eqref{c0TFI0}
  (dashed red line).}
\label{fig.all}
\end{figure}

We shall next show that the previous analysis essentially holds for the HS and FI chains as well.
To begin with, note that from Eqs.~\eqref{f0T}--\eqref{u0T} it follows that
\begin{equation}\label{u0}
f(0,0)=u(0,0)=\frac {K\vep_0}4\,(1-\sgn K)\equiv u_0\,,
\end{equation}
and thus
\begin{align}
  f(0,T)-u_0&=-\frac1\be\int_0^1\log\Bigl(1+\e^{-\frac{|K|\be}2\vep(x)}\Bigr)\diff x\,,\label{f00}\\[1mm]
  u(0,T)-u_0&=\frac{|K|}2\int_0^1\frac{\vep(x)}{1+\e^{\frac{|K|\be}2\vep(x)}}\,\diff x\label{u00}\,.
\end{align}
On the other hand, the free energy $f_\De$ and the internal energy $u_\De$ of a two-level system
with gap $2\De$, when normalized so that $u_\De(0)=f_\De(0)=0$, are given by
\begin{equation}
  f_\De(T)=-\frac1\be\log(1+\e^{-2\be\De})\,,\qquad
  u_\De(T)=\frac{2\De}{1+\e^{2\be\De}}\,.\label{fuDe}
\end{equation}
From the latter equation and Eqs.~\eqref{sDe}--\eqref{cDe}, it follows that the thermodynamic
functions of a two-level system with gap $2\De$ are obtained by taking $\vep(x)$ in
Eqs.~\eqref{s0T}--\eqref{c0T} and~\eqref{f00}--\eqref{u00} as the constant
\begin{equation}
\vep_\De=\frac{4\De}{|K|}\,.
\end{equation}
This observation suggests that, if the ratio $\De/|K|$ is chosen appropriately, the thermodynamic
functions of a two-level system with gap $2\De$ and those of a spin chain of HS type with coupling
$K$ (and zero magnetic field) should behave in a qualitatively similar fashion. In order to
establish the relation between $\De$ and $K$, we note that in the high-temperature limit one has
\[
\lim_{T\to\infty}\Big[u(0,T)-u_0\Big]=\frac{|K|\vep_0}4\,,\qquad
\lim_{T\to\infty}u_\De(T)=\De\,.
\]
For both limits to coincide, we must have
\[
\frac{\De}{|K|}=\frac{\vep_0}4\,,
\]
i.e, $\vep_\De$ should be equal to the average of $\vep(x)$ over the interval $[0,1]$. In
particular, in the case of the PF chain we have $\vep_0=1/2$ (cf.~Eq.~\eqref{vep0}), which explains
the choice $\De=|K|/8$ in Fig.~\ref{fig.PF}. As an example, we have plotted in Fig.~\ref{fig.all}
(left) the specific heat per site of the PF, HS, and FI chains (with $\ga=0,1$ in the latter case)
compared to its counterpart for a two-level system with half-gap $\De=|K|/8$. Here $K$ denotes the
coupling of the PF chain and, following the previous analysis, we have fixed the corresponding
couplings $K_i$ of the other chains by the requirement $K_i\int_0^1\vep_i(x)\,\diff
x=K\int_0^1x\,\diff x=K/2$.

We shall next derive the low temperature behavior of the thermodynamic functions of the HS and FI
chains in the absence of a magnetic field. Our strategy shall be to obtain an asymptotic formula
for the specific heat~\eqref{c0T}, which we shall then integrate to obtain similar formulas for the
remaining thermodynamic functions.

Consider first the HS chain. Since in this case $\vep(x)=x(1-x)$ is symmetric about $x=1/2$, we
can write Eq.~\eqref{c0T} as
\[
\frac{c(0,T)}{\kB}=2\la^2\int_0^{1/2}\frac{\vep^2(x)\,\e^{-\la\vep(x)}}{(1+\e^{-\la\vep(x)})^2}\,\diff
x\,,
\]
where $\la\equiv|K|\be/2$ is a large parameter. Performing the change of variables
$t=\la\,\vep(x)$ we immediately obtain
\[
\frac{c(0,T)}{\kB}=\frac2\la\int_0^{\la/4}\frac{t^2\,\e^{-t}}{(1+\e^{-t})^2}\,\frac{\diff
  t}{\sqrt{1-\frac{4t}\la}}
= \frac2\la\int_0^{\infty}\frac{t^2\,\e^{-t}}{(1+\e^{-t})^2}\,\diff t+\Or(\la^{-2})
\]
(see~\ref{app.error} for the details). The last integral is easily computed:
\[
\int_0^{\infty}\frac{t^2\,\e^{-t}}{(1+\e^{-t})^2}=
\sum_{n=1}^\infty(-1)^{n+1}n\int_0^\infty t^2\,\e^{-nt}\,\diff t=
2\sum_{n=1}^\infty\frac{(-1)^{n+1}}{n^2}=-2\Li_2(-1)=\frac{\pi^2}6
\]
(cf.~Eq.~\eqref{dilogid}), which combined with the previous equation yields
\begin{equation}
  \label{c0THS}
  \frac{c(0,T)}{\kB}=\frac{2\pi^2}{3|K|\be}+O(T^{2})\,.
\end{equation}
Note that the leading term in this formula is twice that in Eq.~\eqref{c0TPF} for the PF chain, due
to the fact that in the present case the leading contribution to the integral~\eqref{c0T} comes
from both endpoints. Apart from this inessential difference, the specific heat at low temperature
behaves as its counterpart for the PF chain, i.e., it increases linearly with the temperature.

Consider next the FI chain. If $\ga\ne0$, proceeding as before we obtain
\begin{align}
  \frac{c(0,T)}{\kB}&=\la^2\int_0^1\frac{\vep^2(x)\,\e^{-\la\vep(x)}}{(1+\e^{-\la\vep(x)})^2}\,\diff
  x= \frac1\la\int_0^{(\ga+1)\la}\frac{t^2\,\e^{-t}}{(1+\e^{-t})^2}\,\frac{\diff
    t}{\sqrt{\ga^2+\frac{4t}\la}}\nonumber\\
  & = \frac1{\ga\la}\int_0^{\infty}\frac{t^2\,\e^{-t}}{(1+\e^{-t})^2}\,\diff
  t+\Or(\la^{-2})=\frac{\pi^2}{3\ga|K|\be}+O(T^{2})\,.
  \label{c0TFIne0}
\end{align}
Again, in the low temperature range the specific heat increases linearly with the temperature.

Finally, for the FI chain with $\ga=0$ we have
\begin{align*}
\frac{c(0,T)}{\kB}&=\la^2\int_0^1\frac{x^4\,\e^{-\la x^2}}{(1+\e^{-\la x^2})^2}\,\diff x=
\frac1{2\sqrt\la}\int_0^{\la}\frac{t^{3/2}\,\e^{-t}}{(1+\e^{-t})^2}\,\diff t\\
&= \frac1{2\sqrt\la}\int_0^\infty\frac{t^{3/2}\,\e^{-t}}{(1+\e^{-t})^2}\,\diff t+
\Or(\la\,\e^{-\la})\,,
\end{align*}
where the last integral can again be exactly evaluated:
\begin{align*}
\int_0^\infty\frac{t^{3/2}\,\e^{-t}}{(1+\e^{-t})^2}\,\diff t&=\sum_{n=1}^\infty(-1)^{n+1}
n\int_0^{\infty}t^{3/2}\,\e^{-nt}\,\diff t\\
&=\sum_{n=1}^\infty\frac{(-1)^{n+1}}{n^{3/2}}
\int_0^{\infty}t^{3/2}\,\e^{-t}\,\diff t=\eta(3/2)\,\Ga(5/2) =\frac{3\sqrt\pi}4\,\eta(3/2)\,,
\end{align*}
where
\[
\eta(z)=\sum_{n=1}^\infty\frac{(-1)^{n+1}}{n^z}
\]
is Dirichlet's eta function. Using the elementary relation
\[
\eta(z)=\left(1-2^{1-z}\right)\kern0pt\ze(z)\,,
\]
where $\ze(z)$ is Riemann's zeta function, we finally obtain
\[
\int_0^\infty\frac{t^{3/2}\,\e^{-t}}{(1+\e^{-t})^2}\,\diff
t=\frac{3\sqrt\pi}{4}\,\left(1-\frac1{\sqrt2}\right)\,\ze(3/2)\,,
\]
and hence
\begin{equation}
  \label{c0TFI0}
  \frac{c(0,T)}{\kB}=\frac38\,(\sqrt 2-1)\,\ze(3/2)\,\sqrt{\frac\pi{|K|\be}}
  +\Or(\be\,\e^{-|K|\be/2})\,.
\end{equation}
In contrast with the previous cases, the specific heat $c(0,T)$ has now an infinite first
derivative at $T=0$; see Fig.~\ref{fig.all} (right). Hence the FI chain with $\ga=B=0$ exhibits a
second-order phase transition at zero temperature.

Since asymptotic expansions can in general be integrated termwise to yield valid asymptotic
expansions~\cite{Er56}, integrating the asymptotic formulas~\eqref{c0TPF} and
\eqref{c0THS}--\eqref{c0TFI0} we can readily derive the low temperature behavior of the remaining
thermodynamic functions of all spin chains of HS type. Thus, for the FI chain with $\ga=0$ we have
\begin{align}
  \label{fFI0}
  f(0,T)&=u_0-\frac{4\eta}3\,\be^{-3/2}+\Or(\be^{-2}\e^{-|K|\be/2})\\[1mm]
  \label{uFI0}
  u(0,T)&=u_0+\frac{2\eta}3\,\be^{-3/2}+\Or(\be^{-1}\e^{-|K|\be/2})\\[1mm]
  \label{sFI0}
  \frac{s(0,T)}{\kB}&=2\eta\,\be^{-1/2}+\Or(\e^{-|K|\be/2})\,,
\end{align}
where we have set
\[
\eta=\frac38\,(\sqrt 2-1)\,\ze(3/2)\,\sqrt{\frac\pi{|K|}}\simeq0.719227\,|K|^{-1/2}\,.
\]
For all other chains of HS type, the analogous results can be concisely summarized as follows:
\begin{align}
  \label{thermoFI0}
  f(0,T)&=u_0-\frac{\eta\,\pi^2}{6|K|\be^2}+\Or(T^3)\\[1mm]
  u(0,T)&=u_0+\frac{\eta\,\pi^2}{6|K|\be^2}+\Or(T^3)\\[1mm]
  \frac{s(0,T)}{\kB}&=\frac{\eta\,\pi^2}{3|K|\be}+\Or(T^2)\,,
\end{align}
where now
\[
\eta=\begin{cases}  2\,,&\HS\\
  1\,,&\PF\\
  1/\ga\,,&\FI{} with \ga\ne0.
\end{cases}
\]

\section{The zero temperature limit}\label{sec.zeroT}

In this section we shall analyze the low temperature behavior of the thermodynamic functions of a
spin chain of HS type in the presence of a nonzero magnetic field. In the ferromagnetic case
($K>0$), we already know from the previous section that there must be a phase transition at
$T=B=0$. This fact is also apparent from Eq.~\eqref{M} for the magnetization, which for $K>0$
implies that
\begin{equation}\label{muT0ferro}
\lim_{T\to0}\mu(B,T)=\sgn B\,.
\end{equation}
The latter result can be corroborated by a computation of the zero-temperature magnetization per
site directly from its definition. Indeed, from Eqs.~\eqref{EbsiTL}-\eqref{epi} it follows that
\begin{equation}\label{muB0}
\mu(B,0)=\lim_{N\to\infty}\frac1N\left\langle\sum_{i=1}^N\si_i\right\rangle_{\!\!\!\rm g}\,,
\end{equation}
where $\langle\,\cdot\,\rangle_{\rm g}$ denotes the average over all ground states. When $B\ne0$
and $K>0$ the ground state is obtained by taking $\si_i=\sgn B$ for all $i$ in Eq.~\eqref{Ebsi0},
and is therefore nondegenerate. Hence for $B\ne0$ we have
\[
\mu(B,0)=\lim_{N\to\infty}\frac1N\,\cdot N\sgn B=\sgn B\,,
\]
as before. On the other hand, when $B=0$ the ground states are obtained from sequences of the form
$\bsi=(1,\dots,1,-1,\dots,-1)$. The average of $\sum_{i=1}^N\si_i$ over such sequences clearly
vanishes, since the contributions of the sequences with $k$ $1$'s and $N-k$ $1$'s cancel each
other. Thus $\mu(0,0)=0$, again in agreement with Eq.~\eqref{muT0ferro}.

In the antiferromagnetic case, it can be shown from Eq.~\eqref{M} that there is a phase transition
at zero temperature and magnetic field $B=\pm\Bs$, where the \emph{saturation field} $\Bs$ is
given by
\begin{equation}\label{Bs}
  \Bs=\frac12\,{|K|}\max_{x\in[0,1]}\vep(x)=
  \begin{cases}
    \frac{|K|}8\,,&\HS\vrule width0pt depth12pt height 17pt\\
    \frac{|K|}2\,,&\PF\vrule width0pt depth12pt\\
    \frac{|K|}2\,(\ga+1)\,,&\FI.\vrule width0pt depth9pt
  \end{cases}
\end{equation}
Indeed, when $K<0$ we can rewrite Eq.~\eqref{M} as
\[
|\mu(B,T)|=\ka\int_0^{1/\ka}\Big[1+4(1-\e^{-2\be|B|})^{-2}\e^{\la\vp(x)}\Big]^{-1/2}\,\diff x\,,
\]
where $\la=|K|\be\to\infty$,
\begin{equation}\label{kappa}
\ka=\begin{cases}1\,,&\text{for the PF and FI chains}\vrule depth9pt width0pt\\2\,,&\HS,
\end{cases}
\end{equation}
and
\begin{equation}
  \label{vp}
  \vp(x)=\vep(x)-\frac{2|B|}{|K|}\equiv\vep(x)-b\,.
\end{equation}
Note that the value $\ka=2$ for the HS chain is due to the symmetry of $\vep(x)=x(1-x)$ about
$x=1/2$ in this case. If $|B|>\Bs$ then
$b>\max\limits_{x\in[0,1]}\vep(x)=\max\limits_{x\in[0,\ka]}\vep(x)$, so that for all $x\in[0,\ka]$
we have $\vp(x)<0$ and $\e^{\la\vp(x)}\to0$ as $\la\to\infty$. Hence
\begin{equation}\label{Msat}
|B|>\Bs\quad\Longrightarrow\quad\lim_{T\to0}|\mu(B,T)|=\ka\int_0^{1/\ka}\diff x=1\,,
\end{equation}
i.e, for $|B|>\Bs$ the magnetization per site saturates. On the other hand, if $|B|\le\Bs$ we have
$\vp(x)<0$ for $0\le x<x_0$ and $\vp(x)>0$ for $x_0<x\le\ka$, where $x_0$ denotes the
unique root in $[0,\ka]$ of the equation $\vp(x)=0$, i.e,
\begin{equation}\label{x0}
\vep(x_0)=b\,,\qquad {\rm with}\quad x_0\in[0,\ka]\,. 
\end{equation}
Thus in this case we have
\[
\lim_{\la\to\infty}\e^{\la\vp(x)}=\begin{cases}0\,,&\text{for } 0\le x<x_0\vrule width0pt
  depth9pt\\\infty\,,&\text{for }x_0< x\le\ka\,,
\end{cases}
\]
and therefore
\begin{equation}\label{Munsat}
|B|\le\Bs\quad\Longrightarrow\quad\lim_{T\to0}|\mu(B,T)|=\ka\int_0^{x_0}\diff x=\ka\,x_0\,.
\end{equation}
Note that the RHS in both Eqs.~\eqref{Msat} and~\eqref{Munsat} tends to $1$ as $|B|\to\Bs$, since
clearly $x_0$ tends to $1/\ka$ as $|B|\to\Bs$.

For the HS chain we have $\vep(x)=x(1-x)$, and hence
\begin{equation}\label{x0HS}
x_0=\frac12-\sqrt{\frac14-b}\equiv \frac12-\sqrt{\frac14-\frac{2|B|}{|K|}}\,.
\end{equation}
Equations~$\eqref{Bs}$ and \eqref{Msat}-\eqref{Munsat} (with $\ka=2$) immediately yield
\begin{equation}\label{muAFHS}
  \lim_{T\to0}\mu(B,T)=\begin{cases}\sgn B\left(1-\sqrt{1-\case{8|B|}{|K|}}\right)\,,
    &|B|\le\frac{|K|}8\vrule width0pt depth12pt \\
    \sgn B\,,&|B|>\frac{|K|}8\,,
  \end{cases}
\end{equation}
which was first derived by Haldane using the spinon gas formalism~\cite{Ha91}.
Likewise, for the PF chain we have $\ka=1$, $\vep(x)=x$ and thus $x_0=b\equiv 2|B|/|K|$, so that
\begin{equation}\label{muAFPF}
\lim_{T\to0}\mu(B,T)=\begin{cases}\case{2B}{|K|}\,,&|B|\le\frac{|K|}2\vrule width0pt depth12pt \\
\sgn B\,,&|B|>\frac{|K|}2\,.
\end{cases}
\end{equation}
Finally, in the case of the FI chain $\ka=1$, $\vep(x)=x(\ga+x)$,
\begin{equation}\label{x0FI}
x_0=-\frac\ga2+\sqrt{\frac{\ga^2}4+b}\equiv-\frac\ga2+\sqrt{\frac{\ga^2}4+\frac{2|B|}{|K|}}\,,
\end{equation}
and thus
\begin{equation}\label{muAFFI}
\lim_{T\to0}\mu(B,T)=\begin{cases}\case12\sgn B\,\left(\sqrt{\ga^2+\case{8|B|}{|K|}}-\ga\right)\,,
  &|B|\le\frac{|K|}2\,(\ga+1)\vrule width0pt depth12pt\\
  \sgn B\,,&|B|>\frac{|K|}2\,(\ga+1)\,,
\end{cases}
\end{equation}
in agreement with the result in Ref.~\cite{FI94}.

\begin{figure}[h]
  \includegraphics[height=3.9cm]{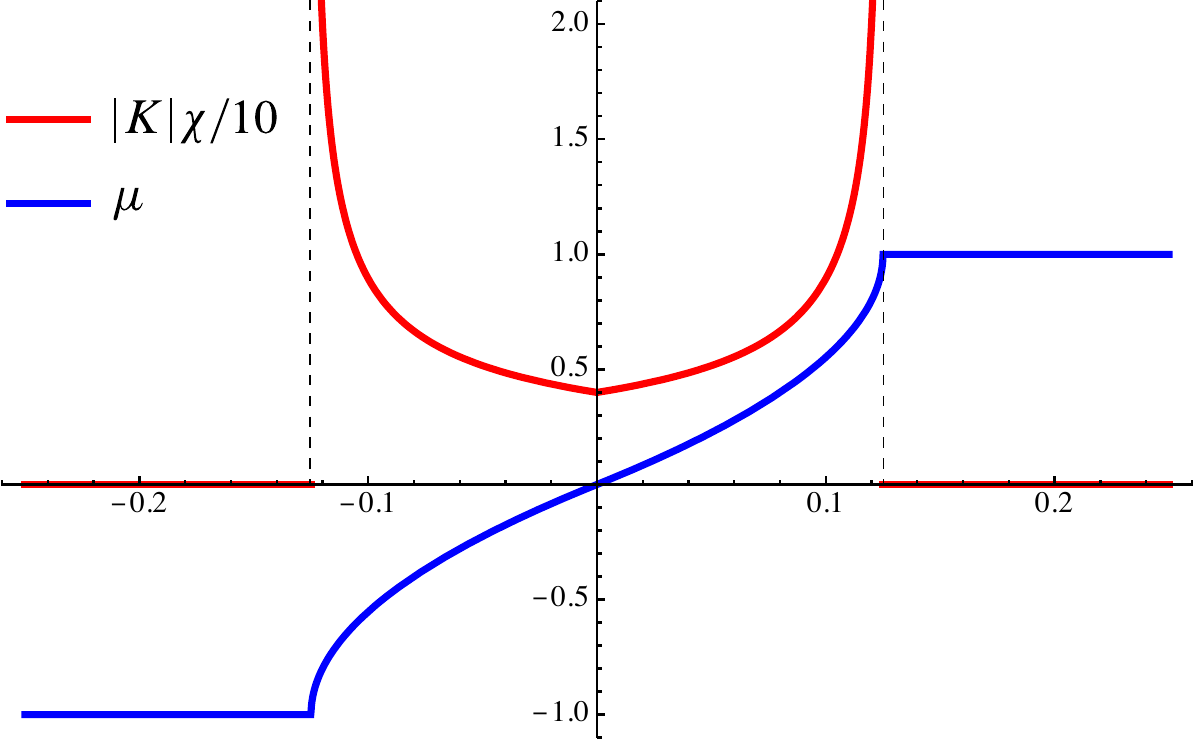}\hfill 
\includegraphics[height=3.9cm]{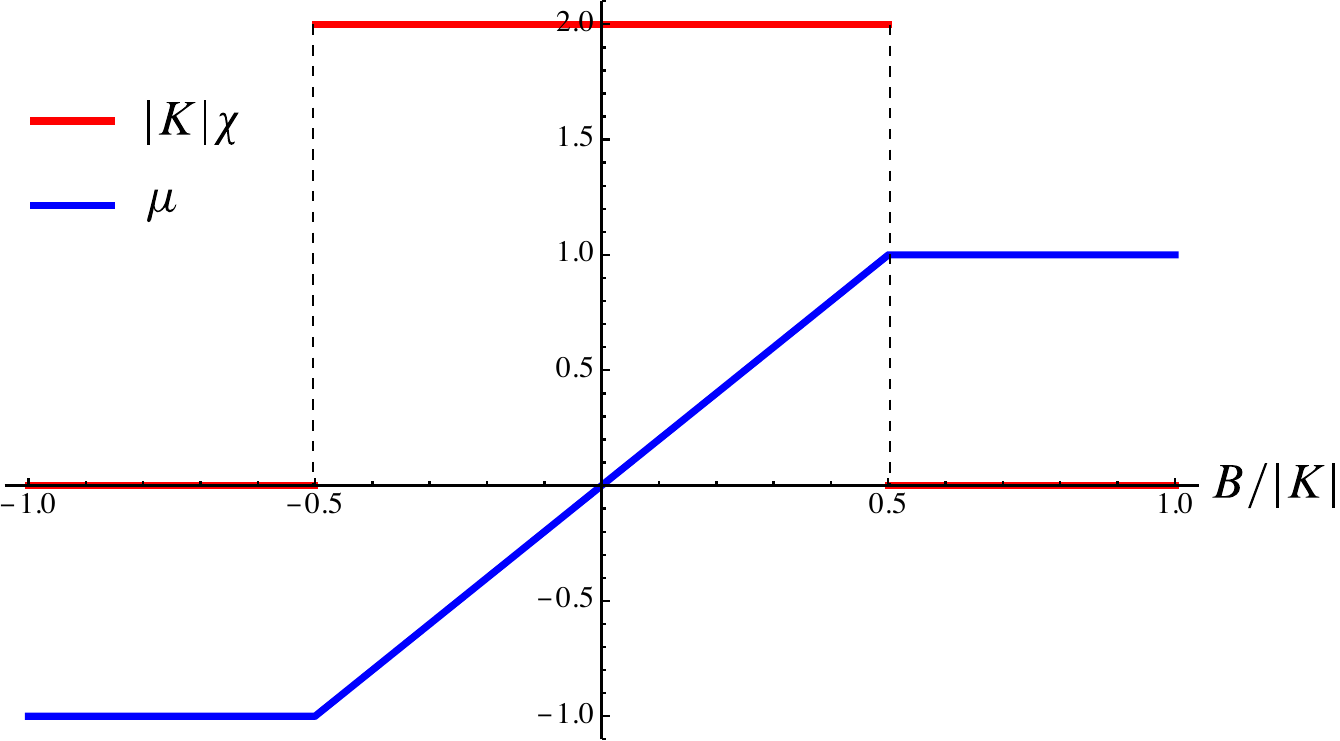}\\[1cm]
\null\kern6pt\includegraphics[height=3.9cm]{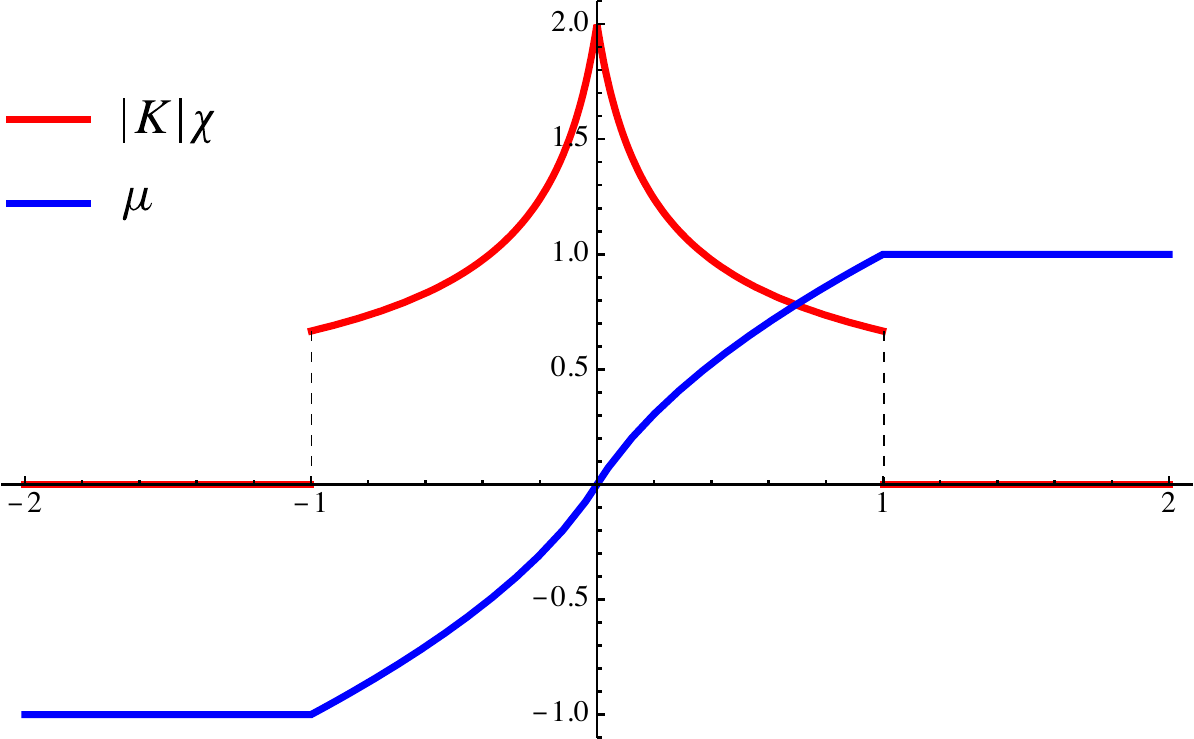}\hfill
\includegraphics[height=3.9cm]{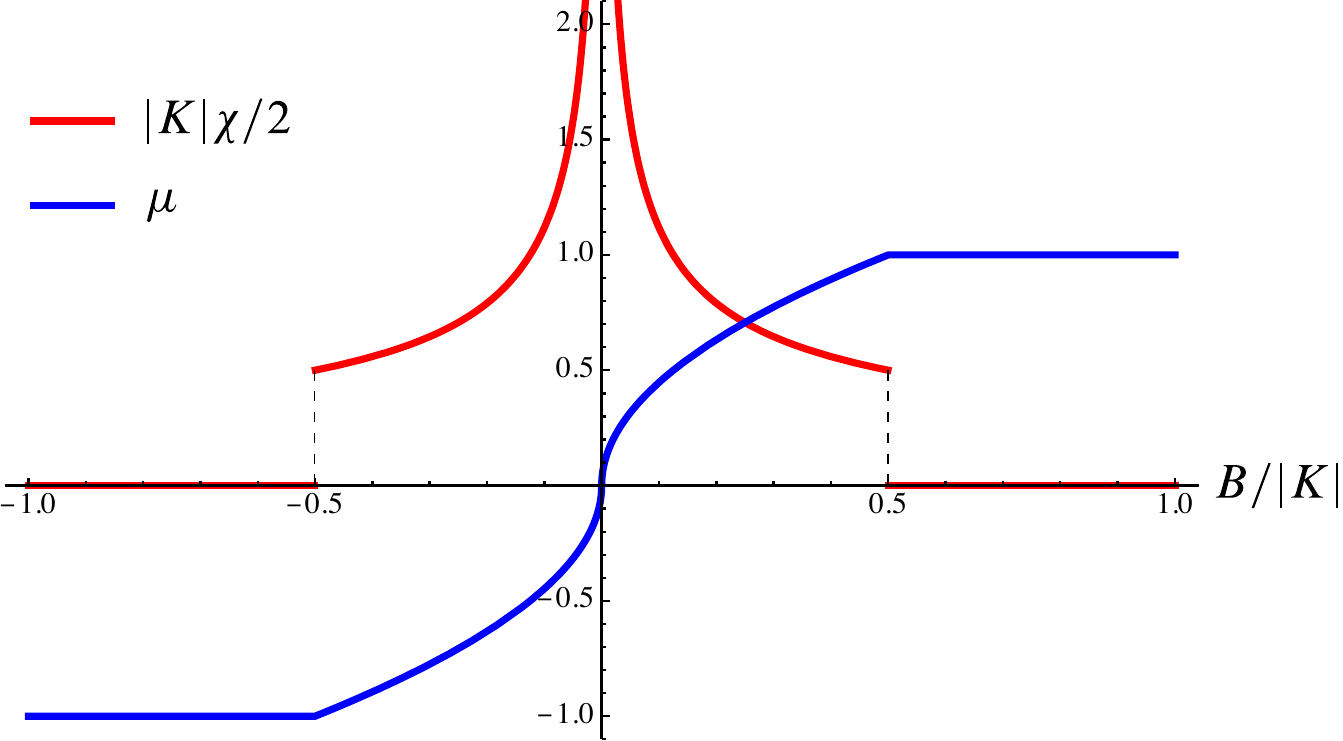}
\caption{Top to bottom and left to right: zero-temperature magnetization and susceptibility per
  site of the HS, PF, and FI chains (with $\ga=1$ and $\ga=0$ for the latter chain) as functions
  of the external magnetic field $B$.}
\label{fig.AFpt}
\end{figure}

From Eqs.~\eqref{muAFHS}--\eqref{muAFFI} it is apparent that, although the magnetization per site
is continuous at $B=\pm\Bs$, the susceptibility $\chi=\frac{\partial\mu}{\partial B}$ has a
discontinuity at these points (cf.~Fig.~\ref{fig.AFpt}). Hence, in the antiferromagnetic case all
three chains of HS type present a second-order phase transition at $T=0$ and $B=\pm\Bs$, where the
saturation field $\Bs$ is explicitly given in Eq.~\eqref{Bs}. The precise behavior of the
zero-temperature magnetization and susceptibility for $|B|\le\Bs$ is, however, quite different for
each type of chain. Indeed, in the case of the HS chain the susceptibility is an increasing
function of $B$ for $0\le B<\Bs$, and diverges as $(\Bs-|B|)^{-1/2}$ when $B\to\pm\Bs\mp$. On the
other hand, for both the PF and FI chains the susceptibility has only a jump discontinuity at the
critical points $B=\pm\Bs$. When $0\le B<\Bs$ the susceptibility is constant for the PF chain, while
for the FI chain it decreases monotonically. Finally, it should be noted that the first derivative
of the zero-temperature susceptibility has a jump discontinuity at $B=0$ for the FI chain with
$\ga>0$ and the HS chain, while it diverges as $|B|^{-1/2}$ at this point for the FI chain with
$\ga=0$ (cf.~Fig.~\ref{fig.AFpt}).

As in the ferromagnetic case, the zero-temperature magnetization can also be computed directly
taking advantage of the exact knowledge of the spectrum. Indeed, consider first the PF and FI
chains. In this case the term proportional to $K=-|K|$ in Eqs.~\eqref{EbsiTL}-\eqref{epi} is
minimized by the sequence $\bsi_0=(\dots,-1,1,\dots,-1,1,-1,1)$,
characteristic of an antiferromagnetic system, while the term proportional to $B$ is again
minimized by the sequence with all components equal to $\sgn B$. It is therefore clear that the
ground state is obtained by flipping the components of $\bsi_0$ equal to $-\sgn B$ starting from
the \emph{left}, until the resulting decrease $2|B|$ in the term proportional to $B$ in
Eq.~\eqref{EbsiTL} is offset by the increase $|K|\vep_i\equiv|K|\vep(x_i)$ in the other term. Thus
if $|B|>\frac12|K|\max\limits_{1\le i\le N}\vep(x_i)=\Bs$ the ground state is again obtained by
taking $\si_i=\sgn B$ for all $i=1,\dots,N$, and the magnetization attains its saturation value
$\sgn B$. On the other hand, when $|B|\le\Bs$ the ground state is obtained from the sequence
\begin{equation}\label{gsPFFI}
(\overbrace{\vrule width0pt height 10pt\sgn B,\dots,\sgn B}^i,-1,1,-1,1,\dots,-1,1)\,,
\end{equation}
where for large $N$ the index $i$ is (approximately) determined by the condition
\begin{equation}\label{i}
|K|\vep(x_i)\simeq 2|B|\,.
\end{equation}
By Eq.~\eqref{muB0}, the magnetization per site is therefore given\footnote{More rigorously, for
  large $N$ there could be a small number of ground states with corresponding sequences almost
  identical to~\eqref{gsPFFI}, so that the average of $\sum_{i=1}^N\si_i$ over these ground states
  is approximately equal to its value on the latter state.}
\begin{equation}\label{Mi}
  \mu(B,0)=\sgn B\cdot\lim_{N\to\infty}\,\frac iN\equiv\sgn B\cdot\lim_{N\to\infty} x_i\,.
\end{equation}
Taking into account that $\lim\limits_{N\to\infty}x_i=x_0$, where as before $x_0$ is determined by
Eq.~\eqref{x0} (with $\ka=1$), we obtain the following formula for the zero-temperature
magnetization per site for $|B|\le\Bs$:
\[
  \mu(B,0)=\sgn B\cdot x_0\,.
\]
The latter equation clearly agrees with Eq.~\eqref{Munsat} with $\ka=1$. For the HS chain the
analysis is very similar, except that the sequence $\bsi_0$ minimizing the term proportional to
$K$ in Eq.~\eqref{EbsiTL} is now of the form
\[
\bsi_0=(\dots,-1,1,\dots,-1,1,\kern-6pt
\mathrel{\mathop{\kern0pt -1}\limits_{\hphantom{-{}}\mathop{\kern0pt \downarrow}\limits_{%
      \vrule width0pt height7pt\scriptstyle[N/2]}}}
\kern-6pt,1,-1,1\dots,-1,1,\dots)\,.
\]
This is due to the fact that in this case the dispersion relation $\vep(x)$ has a maximum at
$x=1/2$, is symmetric about this point, and is monotonically increasing in the interval $[0,1/2]$.
For the same reason, if $|B|<\Bs$ the ground state is now obtained from a sequence of the
form\footnote{As before, in general there could be a small number of ground states corresponding
  to sequences nearly equal to Eq.~\eqref{gsHS}, but this does not affect the conclusion for the
  reason given in the previous footnote.}
\begin{equation}\label{gsHS}
  (\overbrace{\vrule width0pt height 10pt\sgn B,\dots,\sgn B}^i,
  -1,1,\dots,-1,1,\overbrace{\vrule width0pt height 10pt\sgn B,\dots,\sgn B}^i)\,,
\end{equation}
where $i$ is again approximately determined by Eq.~\eqref{i}. Thus in this case Eq.~\eqref{Mi}
should be replaced by
\[
\mu(B,0)=\sgn B\lim_{N\to\infty}\,\frac{2i}N=2\sgn B\lim_{N\to\infty} x_i=2\sgn B\cdot x_0\,,
\]
which is essentially Eq.~\eqref{Munsat} for $\ka=2$.

We shall next analyze the low-temperature behavior of the thermodynamic functions of the
antiferromagnetic chains in the physically more interesting regime $0<|B|<\Bs$. To this end, we
start by rewriting Eq.~\eqref{f} for the free energy per site as
\begin{equation}\label{fB}
-\be(f(B,T)+|B|)=\int_0^1\log\mathopen{}\left[\frac{1+\e^{-2\be|B|}}2+
  \frac12\,\sqrt{(1-\e^{-2\be|B|})^2+4\,\e^{\la\vp(x)}}\,\right]\diff x,
\end{equation}
where $\la\equiv|K|\be\to\infty$ and $\vp(x)$ is given by Eq.~\eqref{vp}. It is straightforward to
show that the function
\begin{equation}\label{g}
g(x,z)=\log\mathopen{}\left[\frac{1+z}2+
  \frac12\,\sqrt{(1-z)^2+4\,\e^{\la\vp(x)}}\,\right]
\end{equation}
satisfies
\[
0\le\frac{\partial g(x,z)}{\partial z}\le\frac12\,\e^{-\frac\la2\vp(x)}\,,\qquad z\ge0\,.
\]
Integrating this inequality with respect to $z$ over the interval $[0,\e^{-2\be|B|}]$ we 
obtain
\[
0\le g(x,\e^{-2\be|B|})-g(x,0)\le\frac12\,\e^{-\frac12\la\vp(x)}\e^{-2\be|B|}=
\frac12\,\e^{-\frac\la2\vep(x)}\e^{-\be|B|}\le\frac12\,\e^{-\be|B|}\,.
\]
Integrating now with respect to $x$ over $[0,1]$ and taking into account Eqs.~\eqref{fB}-\eqref{g}
we easily arrive at the asymptotic relation
\begin{align}
  -\be(f(B,T)+|B|)&=\int_0^1\log\mathopen{}\left(\frac{1+
      \sqrt{1+4\,\e^{\la\vp(x)}}}2\,\right)\diff x+\Or(\e^{-\be|B|})\nonumber\\
  &=\ka\int_0^{1/\ka}\log\mathopen{}\left(\frac{1+
      \sqrt{1+4\,\e^{\la\vp(x)}}}2\,\right)\diff x+\Or(\e^{-\be|B|})\,,
  \label{fBasymp}
\end{align}
where we have used the notation introduced in Eq.~\eqref{kappa}. Since $\vp(x)$ is negative for
$0\le x<x_0$ and positive for $x_0<x\le\ka$, it is convenient to rewrite the latter equation as
\begin{align}
  -\frac\be\ka\,\bigg(f(B,T)+|B|&+\frac{\ka|K|}2\int_{x_0}^{1/\ka}\vp(x)\,\diff x\bigg)
  =\int_0^{x_0}\log\mathopen{}\left(\frac{1+
      \sqrt{1+4\,\e^{\la\vp(x)}}}2\,\right)\diff x\nonumber\\
  &+\int_{x_0}^{1/\ka}\log\mathopen{}\left(\frac12\,\e^{-\frac\la2\vp(x)}+
    \sqrt{1+\frac14\,\e^{-\la\vp(x)}}\,\right)\diff x+\Or(\e^{-\be|B|})\,.
  \label{fBasymp2}
\end{align}

Consider now the first integral in the RHS of the previous equation. Performing the change of
variable $t=-\la\,\vp(x)$ and proceeding as in the previous section we readily
obtain
\begin{align}
  \int_0^{x_0}\log\mathopen{}\left(\frac{1+\sqrt{1+4\,\e^{\la\vp(x)}}}2\,\right)\diff x&=
  \frac1\la\int_0^{\la b}\log\mathopen{}\left(\frac{1+\sqrt{1+4\e^{-t}}}2\,\right)
  \frac{\diff t}{\vep'(x(t))}\nonumber\\
  &\hspace{-1.5cm}=\frac1{\la\vep'(x_0)}\int_0^{\infty}\log\mathopen{}\left(\frac{1+\sqrt{1+4\e^{-t}}}2\,\right)\diff t
  +\Or(\la^{-2})\,,\label{I1}
\end{align}
where the error term has been estimated as in~\ref{app.error}. Note that if $0<|B|<\Bs$ we must
have $0<x_0<\ka$, and consequently $\vep'(x_0)>0$. Observe also that, in order for the above
estimate to be valid, we must have $\la b\gg1$, or (disregarding an inessential factor of $2$)
\[
\be|B|\gg1\,.
\]
The second integral in Eq.~\eqref{fBasymp2} can be similarly dealt with by the change of variable
$t=\la\vp(x)$:
\begin{align}
  &\int_{x_0}^{1/\ka}\log\mathopen{}\left(\frac12\,\e^{-\frac\la2\vp(x)}+
    \sqrt{1+\frac14\,\e^{-\la\vp(x)}}\,\right)\diff x\nonumber\\
  &\qquad\quad=\frac1\la\int_0^{\la\vp(1/\ka)}\log\mathopen{}\left(\frac12\,\e^{-t/2}
    +\sqrt{1+\frac14\,\e^{-t}}\,\right)
  \frac{\diff t}{\vep'(x(t))}\nonumber\\
  &\qquad\quad=\frac1{\la\vep'(x_0)}\int_0^{\infty}\log\mathopen{}\left(\frac12\,\e^{-t/2}
    +\sqrt{1+\frac14\,\e^{-t}}\,\right)\diff t
  +\Or(\la^{-2})\,.\label{I2}
\end{align}
This formula is correct provided that $\la\vp(1/\ka)=\be(|K|\vep(1/\ka)-2|B|)\gg1$, or
equivalently (since $|K|\vep(1/\ka)=2\Bs$),
\[
\be(\Bs-|B|)\gg1\,.
\]
Combining Eq.~\eqref{fBasymp2} with Eqs.~\eqref{I1}-\eqref{I2} we obtain
\begin{equation}
  \label{fTlow}
  -\frac\be\ka\,\bigg(f(B,T)+|B|+\frac{\ka\,|K|}2\int_{x_0}^{1/\ka}\vp(x)\,\diff x\bigg)
  =\frac{I_1+I_2}{\la\vep'(x_0)}+\Or(\la^{-2})\,,
\end{equation}
where
\begin{equation}\label{hI12}
I_1=\int_0^{\infty}\log\mathopen{}\left(\frac{1+\sqrt{1+4\e^{-t}}}2\,\right)\diff t\,,\quad
I_2=\int_0^{\infty}\log\mathopen{}\left(\frac12\,\e^{-t/2}
  +\sqrt{1+\frac14\,\e^{-t}}\,\right)\diff t
\end{equation}
and we have omitted an exponentially small term $\Or(\e^{-\be|B|})\ll\Or(\la^{-2})$.
Using the value of these integrals in Eqs.~\eqref{I1value}-\eqref{I2value} of~\ref{app.int}
we finally obtain the asymptotic formula
\begin{equation}
  \label{fTlowfinal}
  f(B,T)=-|B|-\frac{\ka\,|K|}2\int_{x_0}^{1/\ka}\vp(x)\,\diff x
  -\frac{\ka\,\pi^2}{6|K|\be^2\vep'(x_0)}+\Or(T^3)\,,
\end{equation}
provided that
\begin{equation}\label{Brange}
\frac1\be\ll|B|\ll \Bs-\frac1\be\,.
\end{equation}

Since Eq.~\eqref{fTlowfinal} is an asymptotic power series expansion in $\be$, it may be
differentiated termwise~\cite{Er56} to yield corresponding expansions for the remaining
thermodynamic quantities as $T\to0$. In the first place, taking into account that $x_0$ and
$\vp(x)$ are independent of the temperature we immediately obtain
\begin{align}
  \label{thermoLT}
  u(B,T)&=-|B|-\frac{\ka\,|K|}2\int_{x_0}^{1/\ka}\vp(x)\,\diff x+
  \frac{\ka\,\pi^2}{6|K|\be^2\vep'(x_0)}+\Or(T^3)\,,\\
  \frac{c(B,T)}{\kB}&=
  \frac{\ka\,\pi^2}{3|K|\be\vep'(x_0)}+\Or(T^2)\,,\\
  \frac{s(B,T)}{\kB}&= \frac{\ka\,\pi^2}{3|K|\be\vep'(x_0)}+\Or(T^2)\,.\label{sLT}
\end{align}
In particular, in the case of the HS chain Eq.~\eqref{sLT} is in agreement with Haldane's original
result~\cite{Ha91}.

\begin{figure}[h]
  \includegraphics[height=4.25cm]{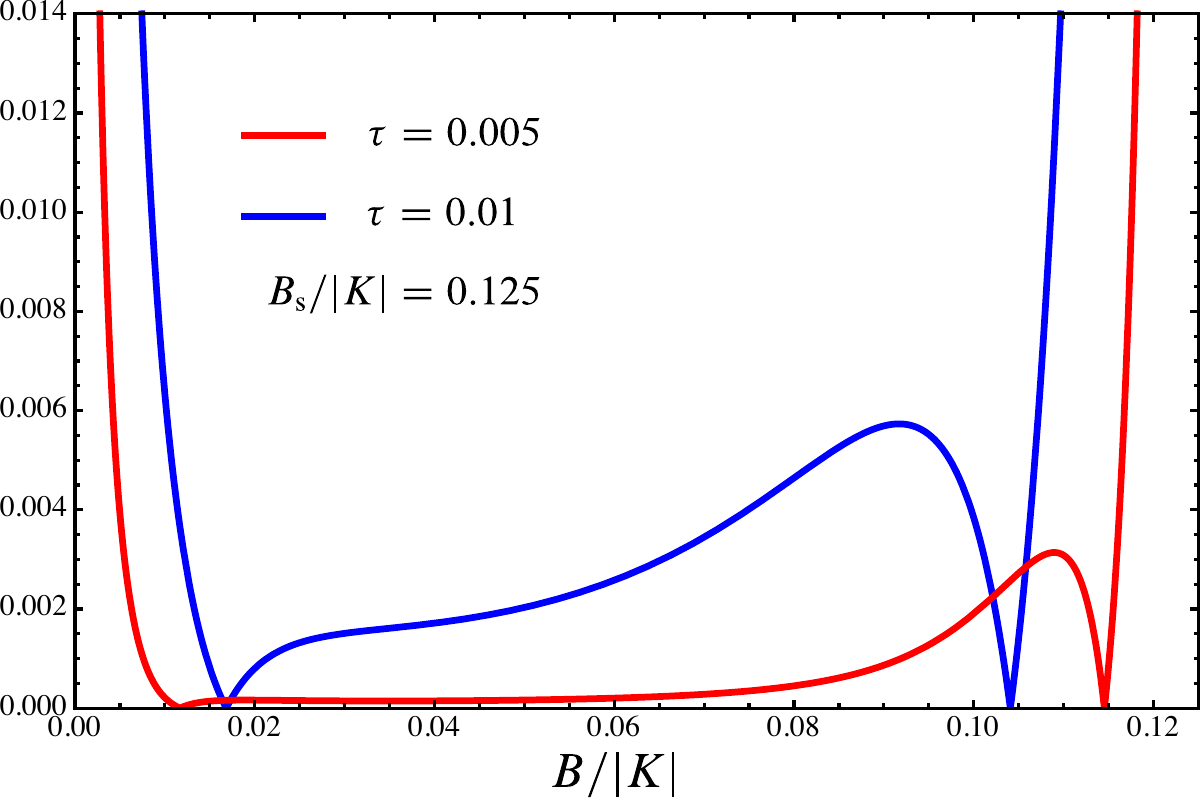}\hfill
  \includegraphics[height=4.17cm]{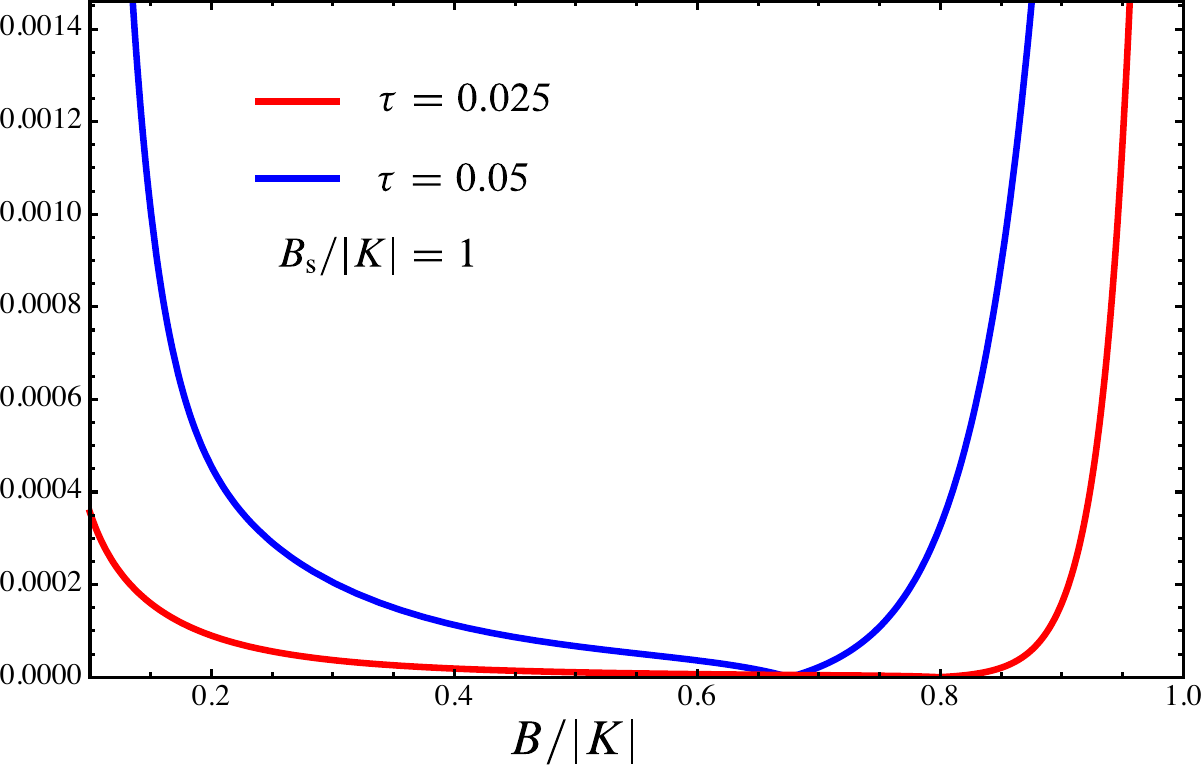}
\caption{Left: absolute value of the relative error of the approximation~\eqref{muLTHS} for HS
the chain for different values of the dimensionless temperature. Right: similar plot for the
approximation~\eqref{muLTFI} for the FI chain with $\ga=1$.}\label{fig.muerrors}
\end{figure}

In order to compute the asymptotic expansion of the magnetization, we observe that, by
Eq.~\eqref{x0},
\[
\frac{\partial x_0}{\partial B}=\frac{2\sgn B}{|K|\vep'(x_0)}\,,
\]
and, since $\vp(x_0)=0$,
\[
\frac{\partial}{\partial B}\int_{x_0}^{1/\ka}\vp(x)\,\diff x
=\int_{x_0}^{1/\ka}\frac{\partial\vp(x)}{\partial B}\,\diff x
=\frac{2\sgn B}{|K|}\left(x_0-\frac1\ka\right)\,.
\]
We thus have
\begin{equation}
  \label{muLT}
  |\mu(B,T)|=
  \ka\,x_0-
  \frac{\ka\,\pi^2\vep''(x_0)}{3K^2\be^2\vep'(x_0)^3}+O(T^3)\,.
\end{equation}
Note, in particular, that the leading term in the latter expansion coincides with
Eq.~\eqref{Munsat}. Using the values of $x_0$ and $\ka$ in Eqs.~\eqref{x0HS}-\eqref{x0FI} and
\eqref{kappa} we easily obtain
\begin{equation}\label{muLTHS}
|\mu(B,T)|=1-\sqrt{1-\case{8|B|}{|K|}}
+\frac{4\,\pi^2}{3K^2\be^2}\left(1-\case{8|B|}{|K|}\right)^{\!-3/2}+\Or(T^3)\,,
\end{equation}
for the HS chain, and
\begin{equation}\label{muLTFI}
|\mu(B,T)|=
\frac12\,\sqrt{\ga^2+\case{8|B|}{|K|}}-\frac\ga2
-\frac{2\,\pi^2}{3K^2\be^2}\left(\ga^2+\case{8|B|}{|K|}\right)^{\!-3/2}+\Or(T^3)\,,
\end{equation}
for the FI chain. In the case of the PF chain the $\Or(T^2)$ term in Eq.~\eqref{muLT} vanishes. In
fact, in this case Eq.~\eqref{MPF} more directly yields
\begin{equation}\label{muLTPF}
|\mu(B,T)|=\frac{2|B|}{|K|}-\frac{\e^{-\frac\be2(|K|-2|B|)}}{|K|\be}+
\Or\left(\frac1\be\,\e^{-\be\min\left(\frac{|K|}2+|B|,\,|K|-2|B|\right)}\right).
\end{equation}
The approximations~\eqref{muLTHS}--\eqref{muLTPF} are in excellent agreement with the exact
result~\eqref{M} (or~\eqref{MPF}, for the PF chain) provided that the magnetic field is in the
range~\eqref{Brange}; see, e.g., Fig.~\ref{fig.muerrors} for the HS and FI chains.

\section{Connection with the one-dimensional Ising model}\label{sec.Ising}

We have seen in Section~\ref{sec.vms} that the chains \eqref{cH2}-\eqref{Jijs} with spin $1/2$ are
isospectral to the classical inhomogeneous Ising model in an external (also inhomogeneous)
magnetic field defined by Eqs.~\eqref{Ebsi} and~\eqref{bidef}. In this section we shall exploit
this fact to study the connection between the above chains and the \emph{standard} (homogeneous)
Ising model, whose Hamiltonian we shall conveniently take as
\begin{equation}
\cH_{J}=J\sum_{i=1}^N(1-S_iS_{i+1})-B\sum_{i=1}^N S_i\,,\qquad S_{N+1}\equiv S_1\,.\label{Ising}
\end{equation}
With this normalization, the free energy per site in the thermodynamic limit is given
by~\cite{Bax82}
\begin{equation}\label{fJ}
f_{J}(T,B)=-\frac1\be\log\Bigl(\cosh(\be B)+\sqrt{\sinh(\be B)^2+\e^{-4\be J}}\,\Bigr)\,,
\end{equation}
which is clearly reminiscent of the analogous Eq.~\eqref{f} for the spin chains of Haldane--Shastry
type. In fact, we can give a heuristic derivation of the latter equation from
Eqs.~\eqref{Ebsi}-\eqref{bidef} and the previous formula for the free energy per site of the Ising
model. To this end, note first of all that the latter equations can be written as
\[
E(\bsi)=\frac K4\sum_{i=1}^{N-1}\vep_i(1-\si_i\si_{i+1})-\sum_{i=1}^N\cB(i)\,\si_i\,,
\]
where
\[
\cB(i)\equiv B+\frac{K}4\,\big(\vep_i-\vep_{i-1}\big)\,.
\]
We saw in Section~\ref{sec.thermo} that when $N\gg1$ the term $\vep_i-\vep_{i-1}$ is
$\Or(N^{-1})$, so that in this limit we have
\[
E(\bsi)\simeq\frac K4\sum_{i=1}^{N-1}\vep_i(1-\si_i\si_{i+1})-B\sum_{i=1}^N\si_i\,.
\]
Proceeding as in the latter section, from the previous equation we can show that when $N\gg 1$ the
partition function of the chains~\eqref{cH2}-\eqref{Jijs} is approximately given by
\[
\cZ\simeq\prod_{i=1}^{N-1}\la_{+}(\textstyle\frac{K\vep_i}4)\,,
\]
where $\la_{+}(J)$ denotes the largest eigenvalue of the transfer matrix the Ising
model~\eqref{Ising} with coupling $J$. Hence, in the thermodynamic limit the free energy per site
of the chains~\eqref{cH2}-\eqref{Jijs} is given by
\begin{equation}\label{fla+}
f(B,T)=-\lim_{N\to\infty}\frac{\log\cZ}{N\be}=-\lim_{N\to\infty}\frac1{N\be}\,\sum_{i=1}^{N-1}
\log\la_{+}(\textstyle\frac{K\vep_i}4)\,.
\end{equation}
On the other hand, when $N\to\infty$ the partition function $\cZ_J$ of the Ising model is related
to the largest eigenvalue $\la_{+}(J)$ of its transfer matrix by~\cite{Bax82}
\[
\cZ_J\simeq\la_+(J)^N, 
\]
so that
\[
f_J(B,T)=-\lim_{N\to\infty}\frac{\log\cZ_J}{N\be}=-\frac1\be\,\log\la_+(J)\,.
\]
Inserting this identity (with $J=\frac{K\vep_i}4$) into Eq.~\eqref{fla+} we obtain
\begin{equation}\label{fff}
f(B,T)=\lim_{N\to\infty}\frac1N\,\sum_{i=1}^{N-1}f_{\frac{K\vep_i}4}(B,T)
=\int_0^1f_{\frac{K\vep(x)}4}(B,T)\,\diff x\,,
\end{equation}
where we have used the fact that $\vep_i=\vep(x_i)\equiv\vep(i/N)$. Equation~\eqref{f}
for the free energy of the spin chains~\eqref{cH2}-\eqref{Jijs} readily follows from the
previous equation and the formula~\eqref{fJ} for $f_J$.

Equation~\eqref{fff} for the free energy of the Haldane--Shastry spin chains admits a statistical
interpretation that we shall discuss next. Indeed, let us first rewrite the latter equation as
\[
f(B,T)=\frac1{\ka}\int_0^\ka f_{\frac{K\vep(x)}4}(B,T)\,\diff x\,,
\]
where $\ka$ is defined in Eq.~\eqref{kappa}. Since the function $\vep(x)$ is monotonically increasing
in the interval $[0,\ka]$, performing the change of variables $J=\frac{K\vep(x)}{4}$ in the latter
integral we immediately obtain
\begin{equation}\label{fav}
f(B,T)=\int_{\min(0,J_0)}^{\max(0,J_0)}\rho(J) f_J(B,T)\,\diff J\,,
\end{equation}
where
\begin{equation}\label{J0}
  J_0=\frac K4\,\vep(\ka)=
  \begin{cases}
    \frac{K}{16}\,,&\HS\vrule width0pt depth12pt height 17pt\\
    \frac{K}4\,,&\PF\vrule width0pt depth12pt\\
    \frac{K}4\,(\ga+1)\,,&\FI\,,\vrule width0pt depth9pt
  \end{cases}
\end{equation}
and
\begin{equation}\label{rho}
\rho(J)=\frac4{\ka |K|\vep'(x)}\,.
\end{equation}
Note that $J_0$ has the sign of $K$, so that the integration range in Eq.~\eqref{fav} is $[0,J_0]$
in the ferromagnetic case and $[J_0,0]$ in the antiferromagnetic one. Furthermore, from
Eq.~\eqref{rho} it immediately follows that
\[
\int_{\min(0,J_0)}^{\max(0,J_0)}\rho(J)\,\diff J=\frac1\ka\int_0^\ka\diff\ka=1\,.
\]
Hence, by Eq.~\eqref{fav}, the free energy of each of the spin chains~\eqref{cH2}-\eqref{Jijs} is the
average of the free energy of an ensemble of standard Ising models weighted by the function
$\rho(J)$ in Eq.~\eqref{rho}. This function, which depends on the chain under consideration through
the dispersion relation $\vep(x)$, is easily computed. Indeed, for the HS chain we have
\[
\vep(x)=x(1-x)=\frac14-\bigg(x-\frac12\bigg)^2\equiv \frac{4J}K\,,
\]
so that
\[
\vep'(x)=1-2x=\sqrt{1-\frac{16J}K}\,.
\]
From Eq.~\eqref{rho} with $\ka=1/2$ we easily obtain
\begin{equation}\label{rhoJHS}
  \rho(J)=\frac{2}{\sqrt{K\big(\frac{K}{16}-J\big)}}=\frac{1}{2\sqrt{J_0(J_0-J)}}\,.
\end{equation}
Similarly, in the case of the PF chain $\vep'(x)=1$ and therefore
\begin{equation}\label{rhoJPF}
\rho(J)=\frac{4}{|K|}=\frac1{|J_0|}
\end{equation}
is constant. Finally, in the case of the FI chain we have
\[
\vep(x)=x(\ga+x)=\bigg(x+\frac\ga2\bigg)^2-\frac{\ga^2}4=\frac{4J}{K}\,,
\]
and therefore
\[
\vep'(x)=2x+\ga=\sqrt{\ga^2+\frac{16J}K}\,.
\]
Hence the weight function $\rho(J)$ is given in this case by
\begin{equation}
  \label{rhoJFI}
  \rho(J)=\frac{4}{\sqrt{K(\ga^2K+16J)}}=\frac{\ga+1}{\sqrt{J_0\big(\ga^2J_0+
    4(\ga+1)J\big)}}\,.
\end{equation}
Introducing the dimensionless variables $|J_0|\rho(J)$ and $j\equiv J/J_0$ (where $j\in[0,1]$ in
the ferromagnetic case and $j\in[-1,0]$ in the antiferromagnetic one) we can rewrite the
previous formulas as
\begin{equation}\label{J0rho}
  |J_0|\rho(J)=
  \begin{cases}
    \frac12(1-j)^{-1/2}\,,&\HS\vrule width0pt depth12pt height 17pt\\
    1\,,&\PF\vrule width0pt depth12pt\\
    (\ga+1)\big[\ga^2+4(\ga+1)j\big]^{-1/2}\,,&\FI\,.\vrule width0pt depth9pt
  \end{cases}
\end{equation}
\begin{figure}[h]
  \centering
  \includegraphics[width=.6\linewidth]{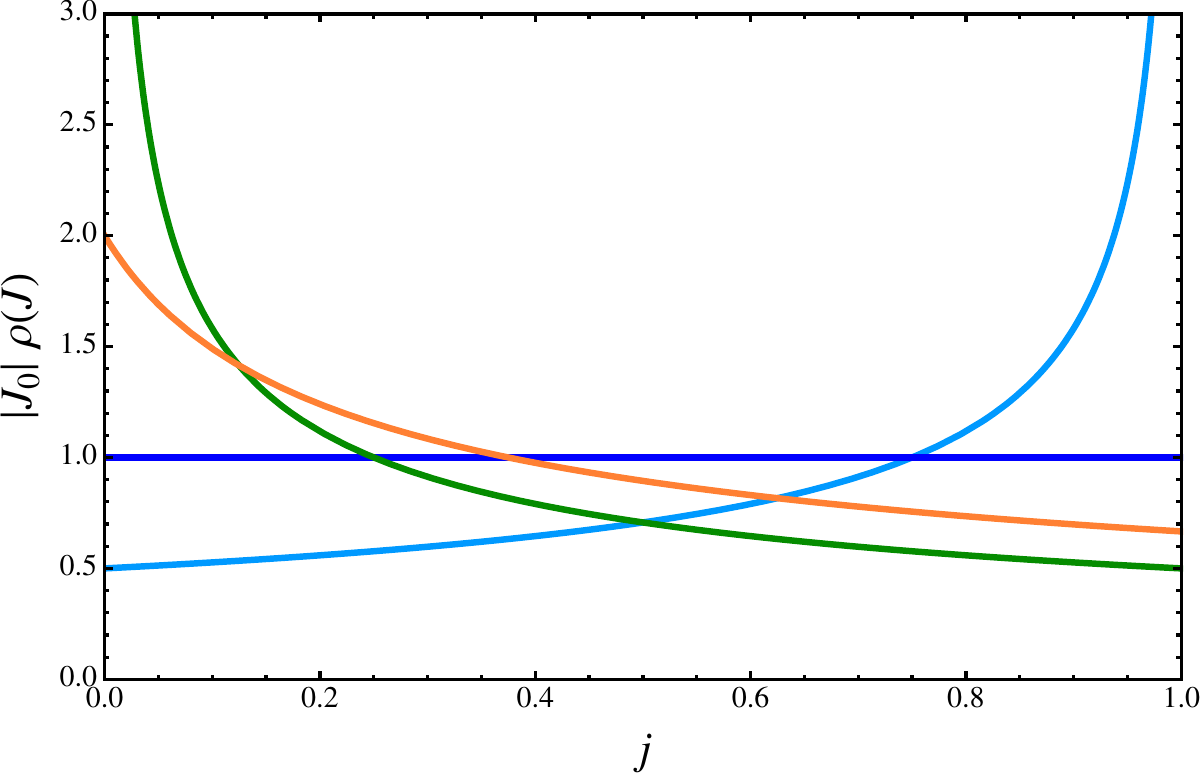}
  \caption{Normalized weight function $|J_0|\rho(J)$ versus dimensionless Ising coupling $j\equiv
    J/J_0$ for the HS (light blue), PF (blue) and FI ($\ga=0$, green; $\ga=1$, orange) chains in the
    ferromagnetic case.}
  \label{fig.rhoJ}
\end{figure}

Equation~\eqref{fav} clearly implies that any thermodynamic quantity of a spin chain of HS type is
the average of this quantity over an ensemble of standard Ising models with respect to the
appropriate weight function $\rho(J)$ in Eqs.~\eqref{rhoJHS}--\eqref{rhoJFI}. In other words, a {\em
  single} spin chain of HS type is thermodynamically equivalent to a suitably weighted {\em
  ensemble} of standard Ising models. {}From this novel point of view, the fundamental difference
between the three types of HS chains~\eqref{cH2}-\eqref{Jijs} is the fact that the weight function
$\rho(J)$ increases with the Ising coupling constant $J$ for the HS chain, is constant for the PF
chain, and decreases with $J$ for the FI chain; cf.~Fig.~\ref{fig.rhoJ}. The study of this weight
function can also uncover some unexpected relations between the different families of chains of HS
type. For instance, from Eq.~\eqref{J0rho} it follows that the normalized densities of the HS chain
and the FI chain with $\ga=0$ are dual, in the sense that they are related by the reflection
$j\mapsto 1-j$ (cf.~Fig.~\ref{fig.rhoJ}).

\section{Conclusions and outlook}\label{sec.concout}

In this paper we derive the thermodynamics of the three families of spin chains of
Haldane--Shastry type~\eqref{cH2}-\eqref{Jijs} in a unified way. The main idea behind our approach
is to exploit the equivalence of these chains with a corresponding family of Ising-like
inhomogeneous vertex models. In the absence of an external magnetic field, this fact had already
been conjectured by Frahm~\cite{Fr93} and by Frahm and Inozemtsev~\cite{FI94}, and was recently
established in a rigorous way in Ref.~\cite{BBH10}. In this paper we generalize the results of the
latter work to the case of an arbitrary magnetic field. Using the equivalence between spin chains
of HS type and inhomogeneous vertex models, in the case of spin $1/2$ we are able to compute the
large $N$ limit of the chains' canonical partition function by means of the transfer matrix
method. We obtain in this way an explicit unified expression of the free energy per site, which we
systematically use for a detailed analysis of the thermodynamic properties. In particular, we
deduce asymptotic formulas for the relevant thermodynamic quantities in both the zero magnetic
field and the zero temperature limits. Finally, we use the explicit formula for the free energy to
show that spin chains of HS type are thermodynamically equivalent to a suitably weighted average
of one-dimensional (\emph{homogeneous}) Ising models.

As remarked throughout the preceding sections, some of the formulas for the thermodynamic
quantities of the HS-type chains derived in this paper have previously appeared in the literature,
especially in the case of the original Haldane--Shastry chain. It should be stressed, however,
that our approach to the thermodynamics of these chains differs from previous related work in two
essential ways. First of all, we treat all three families of spin chains of HS type in a unified
way via their dispersion relation $\vep(x)$ (cf.~\eqref{disprel}). As discussed in the previous
section, this is possible due to the equivalence of these models (in the thermodynamic limit) to
an inhomogeneous Ising model with couplings proportional to the values of the dispersion relation
at the points $x_i\equiv i/N$, $1\le i\le N-1$. On a more technical level, our method for studying
the thermodynamics of spin chains of HS type is simply based on the
evaluation of the canonical partition function in the thermodynamic limit through the standard
transfer matrix approach. We thus bypass the more complex analysis relying on the equivalence of
these models to an ideal gas of spinons used by Haldane to compute the thermodynamic functions of
the original HS chain~\cite{Ha91,BPS93}, which requires the use of the grand canonical ensemble.

The results presented in this paper can be generalized in two different directions.
On the one hand, it should be possible to apply our analysis to general vertex models of the
form~\eqref{EbsiTL}, in which the coupling $\vep_i$ between consecutive spins is an arbitrary
polynomial (or even a $C^1$ function) in the variable $x_i$, $1\le i\le N-1$. These models, which
have been recently studied by Basu-Mallick and collaborators~\cite{BBH10,BB11}, share certain
basic properties with spin chains of HS-type; for instance, their energies are normally
distributed in the thermodynamic limit. On the other hand, it should also be of interest to extend
our results to other types of HS chains, particularly those associated with the $BC_{N}$ and $D_N$
root systems (see, e.g., \cite{EFGR05,BFGR08,BFG09,BFG11}). The first step in this direction would
be to ascertain whether these spin chains are isospectral to suitable vertex models analogous
to~\eqref{EbsiTL}. Work on these and related topics, which is presently going on, shall be
presented in subsequent publications.

\section*{Acknowledgments}
This work was supported in part by the MICINN and the UCM--Banco Santander
under grants no.~FIS2011-22566 and~GR35/10-A-910556.

\appendix

\section{Error term in the asymptotic formulas~\eqref{c0THS}-\eqref{c0TFIne0}}
\label{app.error}

Let us start with the error term in Eq.~\eqref{c0THS}. We must show that
\[
\int_0^{R}\frac{t^2\,\e^{-t}}{(1+\e^{-t})^2}\,\frac{\diff
  t}{\sqrt{1-\frac{t}R}}-\int_0^{\infty}\frac{t^2\,\e^{-t}}{(1+\e^{-t})^2}\,\diff t=\Or(R^{-1})\,,
\]
where we have set $R=\la/4$. Note first of all that the LHS in the previous equation
can be written as
\[
\int_0^{R}\frac{t^2\,\e^{-t}}{(1+\e^{-t})^2}\left[\Big(1-\frac tR\Big)^{-1/2}-1\right]\diff t
-\int_R^{\infty}\frac{t^2\,\e^{-t}}{(1+\e^{-t})^2}\,\diff t\,,
\]
where the last term is easily seen to be $\Or(R^2\,\e^{-R})$. Hence we need only show that
\[
\int_0^{R}\frac{t^2\,\e^{-t}}{(1+\e^{-t})^2}\left[\Big(1-\frac tR\Big)^{\!-1/2}-1\right]\,\diff t
=\Or(R^{-1})\,.
\]
Using the identity
\[
\Big(1-\frac tR\Big)^{-1/2}-1=2R\,\frac{\diff}{\diff t}\left(1-\sqrt{1-\frac tR}-\frac
  t{2R}\right)
\]
and integrating by parts we obtain
\begin{align*}
  \int_0^{R}\frac{t^2\,\e^{-t}}{(1+\e^{-t})^2}&\left[\Big(1-\frac tR\Big)^{-1/2}-1\right]\diff t\\
  &\enspace=2R\int_0^R\left(1-\sqrt{1-\frac tR}-\frac t{2R}\right)g(t)\,\e^{-t}\,\diff t+\Or(R^3\e^{-R})\,,
\end{align*}
where
\[
g(t)=\frac{t^2-2t}{(1+\e^{-t})^2}-\frac{2t^2\e^{-t}}{(1+\e^{-t})^3}\,.
\]
The elementary inequality
\[1-\sqrt{1-s}-\frac s{2}\le \frac{s^2}2\,,\qquad 0\le s\le 1\,,
\]
implies that
\[
0\le\int_0^{R}\frac{t^2\,\e^{-t}}{(1+\e^{-t})^2}\left[\Big(1-\frac tR\Big)^{-1/2}-1\right]\diff t
\le\frac1R\int_0^Rt^2g(t)\e^{-t}\,\diff t+\Or(R^3\e^{-R})=\Or(R^{-1})\,,
\]
since $g(t)=\Or(t^2)$.

Consider next the error term in Eq.~\eqref{c0TFIne0}. We must now show that
\[
\int_0^\infty\frac{t^2\,\e^{-t}}{(1+\e^{-t})^2}\,\diff t-
\int_0^{(\ga+1)\la}\frac{t^2\,\e^{-t}}{(1+\e^{-t})^2}\,\frac{\diff
  t}{\sqrt{1+\frac{4t}{\ga^2\la}}}=\Or(\la^{-1})\,,
\]
or equivalently that
\[
\int_0^{(\ga+1)\la}\frac{t^2\,\e^{-t}}{(1+\e^{-t})^2}\,
\left[1-\left(1+\frac{4t}{\ga^2\la}\right)^{-1/2}\right]\diff t=\Or(\la^{-1})\,,
\]
since
\[
\int_{(\ga+1)\la}^\infty\frac{t^2\,\e^{-t}}{(1+\e^{-t})^2}\,\diff t=\Or(\la^2\e^{-(\ga+1)\la})\,.
\]
Our assertion follows immediately from the elementary inequality
\[
1-(1+s)^{-1/2}\le\frac s2\,,\qquad s>0\,,
\]
which implies that
\[
0\le\int_0^{(\ga+1)\la}\frac{t^2\,\e^{-t}}{(1+\e^{-t})^2}\,
\left[1-\left(1+\frac{4t}{\ga^2\la}\right)^{-1/2}\right]\diff t\le\frac2{\ga^2\la}
\int_0^{(\ga+1)\la}\frac{t^3\,\e^{-t}}{(1+\e^{-t})^2}=\Or(\la^{-1})
\]
on account of the convergent character of the integral
\[
\int_0^\infty\frac{t^3\,\e^{-t}}{(1+\e^{-t})^2}\,\diff t\,.
\]

\section{Computation of the definite integrals~\eqref{hI12}}
\label{app.int}

Consider the definite integral
\[
I(x;\al)=\int_0^{x}\log\mathopen{}\left(\frac{1+\sqrt{1+4\e^{\al t}}}2\,\right)\diff t\,,
\]
with $\al^2=1$. Performing the change of variable
\[
z=\frac12\left(1-\sqrt{1+4\e^{\al t}}\,\right)
\]
and taking into account that
\[
\Li_2\mathopen{}\left(\frac{1-\sqrt5}2\right)=\frac12\,\log^2\mathopen{}\left(\frac{1+\sqrt5}2\right)-\frac{\pi^2}{15}
\]
(cf.~Ref.~\cite{Le81}), we readily obtain
\begin{align}
  \al I(x;\al)&=\int_{\frac12(1-\sqrt5\,)}^{\frac12(1-\sqrt{1+4\e^{\al x}}\,)}
  \left[\frac{\log(1-z)}{z}-\frac{\log(1-z)}{1-z}\right]\diff z\nonumber\\
  &=\frac12\,\log^2\mathopen{}\left(\frac{1+\sqrt{1+4\e^{\al
          x}}}2\right)-\Li_2\mathopen{}\left(\frac{1-\sqrt{1+4\e^{\al x}}\,}2\right)-\frac{\pi^2}{15}\,.
  \label{Ixal}
\end{align}
Hence
\begin{equation}
  \label{I1value}
  I_1=I(\infty;-1)=\frac{\pi^2}{15}\,.
\end{equation}
As to the second integral in Eq.~\eqref{hI12}, we first note that
\[
I_2=\lim_{x\to\infty}\left(I(x;1)-\frac{x^2}4\right).
\]
From eq.~\eqref{Ixal} and the dilogarithm identity~\cite{Le81}
\[
-\Li_2(1-z)=\Li_2\mathopen{}\left(1-\frac1z\right)+\frac12\,\log^2z
\]
it follows that
\[
I(x;1)=\log^2\Biggl(\frac{1+\sqrt{1+4\e^x}}2\Biggr)
+\Li_2\Biggl(1-\frac2{1+\sqrt{1+4\e^x}}\Biggr)-\frac{\pi^2}{15}\,,
\]
and therefore
\begin{align}
  I_2&=\Li_2(1)-\frac{\pi^2}{15}+\lim_{x\to\infty}\left[\log^2\Biggl(\frac{1+\sqrt{1+4\e^x}}2\,\Biggr)
    -\frac{x^2}4\right]\nonumber\\&=\frac{\pi^2}{10}
  +\lim_{x\to\infty}\left[\log^2\Biggl(\frac{1+\sqrt{1+4\e^x}}2\,\Biggr)-\frac{x^2}4\right],
  \label{I2lim}
\end{align}
where we have made use of the equality $\Li_2(1)=\ze(2)=\pi^2/6$ (cf.~Ref.~\cite{Le81}). Since
\[
\log\Biggl(\frac{1+\sqrt{1+4\e^x}}2\,\Biggr)=\frac x2+\Or(\e^{-x/2})\,,
\]
the limit in Eq.~\eqref{I2lim} vanishes, and we finally obtain
\begin{equation}
  I_2=\frac{\pi^2}{10}\,.
  \label{I2value}
\end{equation}


\begin{thebibliography}{49}
\expandafter\ifx\csname natexlab\endcsname\relax\def\natexlab#1{#1}\fi
\providecommand{\bibinfo}[2]{#2}
\ifx\xfnm\relax \def\xfnm[#1]{\unskip,\space#1}\fi
\bibitem[{Polychronakos(2006)}]{Po06}
\bibinfo{author}{A.~P. Polychronakos}, \bibinfo{journal}{J. Phys. A: Math.
  Gen.} \bibinfo{volume}{39} (\bibinfo{year}{2006})
  \bibinfo{pages}{12793--12845}.
\bibitem[{Haldane(1988)}]{Ha88}
\bibinfo{author}{F.~D.~M. Haldane}, \bibinfo{journal}{Phys. Rev. Lett.}
  \bibinfo{volume}{60} (\bibinfo{year}{1988}) \bibinfo{pages}{635--638}.
\bibitem[{Shastry(1988)}]{Sh88}
\bibinfo{author}{B.~S. Shastry}, \bibinfo{journal}{Phys. Rev. Lett.}
  \bibinfo{volume}{60} (\bibinfo{year}{1988}) \bibinfo{pages}{639--642}.
\bibitem[{Polychronakos(1993)}]{Po93}
\bibinfo{author}{A.~P. Polychronakos}, \bibinfo{journal}{Phys. Rev. Lett.}
  \bibinfo{volume}{70} (\bibinfo{year}{1993}) \bibinfo{pages}{2329--2331}.
\bibitem[{Frahm(1993)}]{Fr93}
\bibinfo{author}{H.~Frahm}, \bibinfo{journal}{J. Phys. A: Math. Gen.}
  \bibinfo{volume}{26} (\bibinfo{year}{1993}) \bibinfo{pages}{L473--L479}.
\bibitem[{Frahm and Inozemtsev(1994)}]{FI94}
\bibinfo{author}{H.~Frahm}, \bibinfo{author}{V.~I. Inozemtsev},
  \bibinfo{journal}{J. Phys. A: Math. Gen.} \bibinfo{volume}{27}
  (\bibinfo{year}{1994}) \bibinfo{pages}{L801--L807}.
\bibitem[{Yamamoto and Tsuchiya(1996)}]{YT96}
\bibinfo{author}{T.~Yamamoto}, \bibinfo{author}{O.~Tsuchiya},
  \bibinfo{journal}{J. Phys. A: Math. Gen.} \bibinfo{volume}{29}
  (\bibinfo{year}{1996}) \bibinfo{pages}{3977--3984}.
\bibitem[{Enciso et~al.(2005)Enciso, Finkel, Gonz{\'a}lez-L\'opez, and
  Rodr{{\'\i}}guez}]{EFGR05}
\bibinfo{author}{A.~Enciso}, \bibinfo{author}{F.~Finkel},
  \bibinfo{author}{A.~Gonz{\'a}lez-L\'opez}, \bibinfo{author}{M.~A.
  Rodr{{\'\i}}guez}, \bibinfo{journal}{Nucl. Phys. B} \bibinfo{volume}{707}
  (\bibinfo{year}{2005}) \bibinfo{pages}{553--576}.
\bibitem[{Barba et~al.(2008)Barba, Finkel, Gonz\'alez-L\'opez, and
  Rodr{\'\i}guez}]{BFGR08}
\bibinfo{author}{J.~C. Barba}, \bibinfo{author}{F.~Finkel},
  \bibinfo{author}{A.~Gonz\'alez-L\'opez}, \bibinfo{author}{M.~A.
  Rodr{\'\i}guez}, \bibinfo{journal}{Phys. Rev. B} \bibinfo{volume}{77}
  (\bibinfo{year}{2008}) \bibinfo{pages}{214422(10)}.
\bibitem[{Basu-Mallick et~al.(2009)Basu-Mallick, Finkel, and
  Gonz{\'a}lez-L{\'o}pez}]{BFG09}
\bibinfo{author}{B.~Basu-Mallick}, \bibinfo{author}{F.~Finkel},
  \bibinfo{author}{A.~Gonz{\'a}lez-L{\'o}pez}, \bibinfo{journal}{Nucl. Phys. B}
  \bibinfo{volume}{812} (\bibinfo{year}{2009}) \bibinfo{pages}{402--423}.
\bibitem[{Basu-Mallick et~al.(2011)Basu-Mallick, Finkel, and
  Gonz{\'a}lez-L{\'o}pez}]{BFG11}
\bibinfo{author}{B.~Basu-Mallick}, \bibinfo{author}{F.~Finkel},
  \bibinfo{author}{A.~Gonz{\'a}lez-L{\'o}pez}, \bibinfo{journal}{Nucl. Phys. B}
  \bibinfo{volume}{843} (\bibinfo{year}{2011}) \bibinfo{pages}{505--553}.
\bibitem[{Haldane(1991)}]{Ha91b}
\bibinfo{author}{F.~D.~M. Haldane}, \bibinfo{journal}{Phys. Rev. Lett.}
  \bibinfo{volume}{67} (\bibinfo{year}{1991}) \bibinfo{pages}{937--940}.
\bibitem[{Gebhard and Ruckenstein(1992)}]{GR92}
\bibinfo{author}{F.~Gebhard}, \bibinfo{author}{A.~E. Ruckenstein},
  \bibinfo{journal}{Phys. Rev. Lett.} \bibinfo{volume}{68}
  (\bibinfo{year}{1992}) \bibinfo{pages}{244--247}.
\bibitem[{Fowler and Minahan(1993)}]{FM93}
\bibinfo{author}{M.~Fowler}, \bibinfo{author}{J.~A. Minahan},
  \bibinfo{journal}{Phys. Rev. Lett.} \bibinfo{volume}{70}
  (\bibinfo{year}{1993}) \bibinfo{pages}{2325--2328}.
\bibitem[{Bernard et~al.(1993)Bernard, Gaudin, Haldane, and Pasquier}]{BGHP93}
\bibinfo{author}{D.~Bernard}, \bibinfo{author}{M.~Gaudin},
  \bibinfo{author}{F.~D.~M. Haldane}, \bibinfo{author}{V.~Pasquier},
  \bibinfo{journal}{J. Phys. A: Math. Gen.} \bibinfo{volume}{26}
  (\bibinfo{year}{1993}) \bibinfo{pages}{5219--5236}.
\bibitem[{Ha and Haldane(1992)}]{HH92}
\bibinfo{author}{Z.~N.~C. Ha}, \bibinfo{author}{F.~D.~M. Haldane},
  \bibinfo{journal}{Phys. Rev. B} \bibinfo{volume}{46} (\bibinfo{year}{1992})
  \bibinfo{pages}{9359--9368}.
\bibitem[{Minahan and Polychronakos(1993)}]{MP93}
\bibinfo{author}{J.~A. Minahan}, \bibinfo{author}{A.~P. Polychronakos},
  \bibinfo{journal}{Phys. Lett. B} \bibinfo{volume}{302} (\bibinfo{year}{1993})
  \bibinfo{pages}{265--270}.
\bibitem[{Inozemtsev(1996)}]{In96}
\bibinfo{author}{V.~I. Inozemtsev}, \bibinfo{journal}{Phys. Scr.}
  \bibinfo{volume}{53} (\bibinfo{year}{1996}) \bibinfo{pages}{516--520}.
\bibitem[{Barba et~al.(2008)Barba, Finkel, Gonz\'alez-L\'opez, and
  Rodr{\'\i}guez}]{BFGR08epl}
\bibinfo{author}{J.~C. Barba}, \bibinfo{author}{F.~Finkel},
  \bibinfo{author}{A.~Gonz\'alez-L\'opez}, \bibinfo{author}{M.~A.
  Rodr{\'\i}guez}, \bibinfo{journal}{Europhys. Lett.} \bibinfo{volume}{83}
  (\bibinfo{year}{2008}) \bibinfo{pages}{27005(6)}.
\bibitem[{Barba et~al.(2009)Barba, Finkel, Gonz\'alez-L\'opez, and
  Rodr{\'\i}guez}]{BFGR09power}
\bibinfo{author}{J.~C. Barba}, \bibinfo{author}{F.~Finkel},
  \bibinfo{author}{A.~Gonz\'alez-L\'opez}, \bibinfo{author}{M.~A.
  Rodr{\'\i}guez}, \bibinfo{journal}{Phys. Rev. E} \bibinfo{volume}{80}
  (\bibinfo{year}{2009}) \bibinfo{pages}{047201(4)}.
\bibitem[{Basu-Mallick et~al.(1999)Basu-Mallick, Ujino, and Wadati}]{BUW99}
\bibinfo{author}{B.~Basu-Mallick}, \bibinfo{author}{H.~Ujino},
  \bibinfo{author}{M.~Wadati}, \bibinfo{journal}{J. Phys. Soc. Jpn.}
  \bibinfo{volume}{68} (\bibinfo{year}{1999}) \bibinfo{pages}{3219--3226}.
\bibitem[{Basu-Mallick and Bondyopadhaya(2006)}]{BB06}
\bibinfo{author}{B.~Basu-Mallick}, \bibinfo{author}{N.~Bondyopadhaya},
  \bibinfo{journal}{Nucl. Phys. B} \bibinfo{volume}{757} (\bibinfo{year}{2006})
  \bibinfo{pages}{280--302}.
\bibitem[{Haldane(1991)}]{Ha91}
\bibinfo{author}{F.~D.~M. Haldane}, \bibinfo{journal}{Phys. Rev. Lett.}
  \bibinfo{volume}{66} (\bibinfo{year}{1991}) \bibinfo{pages}{1529--1532}.
\bibitem[{Basu-Mallick et~al.(2008)Basu-Mallick, Bondyopadhaya, and
  Sen}]{BBS08}
\bibinfo{author}{B.~Basu-Mallick}, \bibinfo{author}{N.~Bondyopadhaya},
  \bibinfo{author}{D.~Sen}, \bibinfo{journal}{Nucl. Phys. B}
  \bibinfo{volume}{795} (\bibinfo{year}{2008}) \bibinfo{pages}{596--622}.
\bibitem[{Cirac and Sierra(2010)}]{CS10}
\bibinfo{author}{J.~I. Cirac}, \bibinfo{author}{G.~Sierra},
  \bibinfo{journal}{Phys. Rev. B} \bibinfo{volume}{81} (\bibinfo{year}{2010})
  \bibinfo{pages}{104431(4)}.
\bibitem[{Beisert et~al.(2003)Beisert, Kristjansen, and Staudacher}]{BKS03}
\bibinfo{author}{N.~Beisert}, \bibinfo{author}{C.~Kristjansen},
  \bibinfo{author}{M.~Staudacher}, \bibinfo{journal}{Nucl. Phys. B}
  \bibinfo{volume}{664} (\bibinfo{year}{2003}) \bibinfo{pages}{131--184}.
\bibitem[{Bargheer et~al.(2009)Bargheer, Beisert, and Loebbert}]{BBL09}
\bibinfo{author}{T.~Bargheer}, \bibinfo{author}{N.~Beisert},
  \bibinfo{author}{F.~Loebbert}, \bibinfo{journal}{J. Phys. A: Math. Theor.}
  \bibinfo{volume}{42} (\bibinfo{year}{2009}) \bibinfo{pages}{285205(58)}.
\bibitem[{Greiter(2009)}]{Gr09}
\bibinfo{author}{M.~Greiter}, \bibinfo{journal}{Phys. Rev. B}
  \bibinfo{volume}{79} (\bibinfo{year}{2009}) \bibinfo{pages}{064409(5)}.
\bibitem[{Hikami(1995)}]{Hi95npb}
\bibinfo{author}{K.~Hikami}, \bibinfo{journal}{Nucl. Phys. B}
  \bibinfo{volume}{441} (\bibinfo{year}{1995}) \bibinfo{pages}{530--548}.
\bibitem[{Basu-Mallick(1999)}]{Ba99}
\bibinfo{author}{B.~Basu-Mallick}, \bibinfo{journal}{Nucl. Phys. B}
  \bibinfo{volume}{540} (\bibinfo{year}{1999}) \bibinfo{pages}{679--704}.
\bibitem[{Beisert and Erkal(2008)}]{BE08}
\bibinfo{author}{N.~Beisert}, \bibinfo{author}{D.~Erkal}, \bibinfo{journal}{J.
  Stat. Mech.} \bibinfo{volume}{0803} (\bibinfo{year}{2008})
  \bibinfo{pages}{P03001}.
\bibitem[{Polychronakos(1994)}]{Po94}
\bibinfo{author}{A.~P. Polychronakos}, \bibinfo{journal}{Nucl. Phys. B}
  \bibinfo{volume}{419} (\bibinfo{year}{1994}) \bibinfo{pages}{553--566}.
\bibitem[{Finkel and Gonz{\'a}lez-L\'opez(2005)}]{FG05}
\bibinfo{author}{F.~Finkel}, \bibinfo{author}{A.~Gonz{\'a}lez-L\'opez},
  \bibinfo{journal}{Phys. Rev. B} \bibinfo{volume}{72} (\bibinfo{year}{2005})
  \bibinfo{pages}{174411(6)}.
\bibitem[{Barba et~al.(2010)Barba, Finkel, Gonz\'alez-L\'opez, and
  Rodr{\'\i}guez}]{BFGR10}
\bibinfo{author}{J.~C. Barba}, \bibinfo{author}{F.~Finkel},
  \bibinfo{author}{A.~Gonz\'alez-L\'opez}, \bibinfo{author}{M.~A.
  Rodr{\'\i}guez}, \bibinfo{journal}{Nucl. Phys. B} \bibinfo{volume}{839}
  (\bibinfo{year}{2010}) \bibinfo{pages}{499--525}.
\bibitem[{Basu-Mallick et~al.(2010)Basu-Mallick, Bondyopadhaya, and
  Hikami}]{BBH10}
\bibinfo{author}{B.~Basu-Mallick}, \bibinfo{author}{N.~Bondyopadhaya},
  \bibinfo{author}{K.~Hikami}, \bibinfo{journal}{SIGMA} \bibinfo{volume}{6}
  (\bibinfo{year}{2010}) \bibinfo{pages}{091--13}.
\bibitem[{Haldane et~al.(1992)Haldane, Ha, Talstra, Bernard, and
  Pasquier}]{HHTBP92}
\bibinfo{author}{F.~D.~M. Haldane}, \bibinfo{author}{Z.~N.~C. Ha},
  \bibinfo{author}{J.~C. Talstra}, \bibinfo{author}{D.~Bernard},
  \bibinfo{author}{V.~Pasquier}, \bibinfo{journal}{Phys. Rev. Lett.}
  \bibinfo{volume}{69} (\bibinfo{year}{1992}) \bibinfo{pages}{2021--2025}.
\bibitem[{Calogero(1971)}]{Ca71}
\bibinfo{author}{F.~Calogero}, \bibinfo{journal}{J. Math. Phys.}
  \bibinfo{volume}{12} (\bibinfo{year}{1971}) \bibinfo{pages}{419--436}.
\bibitem[{Corrigan and Sasaki(2002)}]{CS02}
\bibinfo{author}{E.~Corrigan}, \bibinfo{author}{R.~Sasaki},
  \bibinfo{journal}{J. Phys. A: Math. Gen.} \bibinfo{volume}{35}
  (\bibinfo{year}{2002}) \bibinfo{pages}{7017--7061}.
\bibitem[{Enciso et~al.(2008)Enciso, Finkel, Gonz{\'a}lez-L{\'o}pez, and
  Rodr{\'\i}guez}]{EFGR08jnmp}
\bibinfo{author}{A.~Enciso}, \bibinfo{author}{F.~Finkel},
  \bibinfo{author}{A.~Gonz{\'a}lez-L{\'o}pez}, \bibinfo{author}{M.~A.
  Rodr{\'\i}guez}, \bibinfo{journal}{J. Nonlin. Math. Phys.}
  \bibinfo{volume}{15} (\bibinfo{year}{2008}) \bibinfo{pages}{155--165}.
\bibitem[{Enciso(2009)}]{En09}
\bibinfo{author}{A.~Enciso}, \bibinfo{title}{Spin models of
  {C}alogero--{S}utherland type and associated spin chains},
  \bibinfo{year}{2009}. \bibinfo{note}{{P}h.{D}. {T}hesis, {U}niversidad
  {C}omplutense de {M}adrid (arXiv:0906.1167v1 [math-ph])}.
\bibitem[{Finkel et~al.(2001)Finkel, G\'omez-Ullate, Gonz{\'a}lez-L\'opez,
  Rodr{{\'\i}}guez, and Zhdanov}]{FGGRZ01}
\bibinfo{author}{F.~Finkel}, \bibinfo{author}{D.~G\'omez-Ullate},
  \bibinfo{author}{A.~Gonz{\'a}lez-L\'opez}, \bibinfo{author}{M.~A.
  Rodr{{\'\i}}guez}, \bibinfo{author}{R.~Zhdanov}, \bibinfo{journal}{Commun.
  Math. Phys.} \bibinfo{volume}{221} (\bibinfo{year}{2001})
  \bibinfo{pages}{477--497}.
\bibitem[{Macdonald(1995)}]{Ma95}
\bibinfo{author}{I.~G. Macdonald}, \bibinfo{title}{{S}ymmetric Functions and
  {H}all Polynomials}, \bibinfo{publisher}{Oxford University Press},
  \bibinfo{address}{Oxford}, \bibinfo{year}{1995}.
\bibitem[{Bernard et~al.(1993)Bernard, Pasquier, and Serban}]{BPS93}
\bibinfo{author}{D.~Bernard}, \bibinfo{author}{V.~Pasquier},
  \bibinfo{author}{D.~Serban}, \bibinfo{title}{A one-dimensioal ideal gas of
  spinons, or some exact results on the {XXX} spin chain with long range
  interaction}, \bibinfo{year}{1993}. \bibinfo{note}{ArXiv:hep-th/9311013v1}.
\bibitem[{Lewin(1981)}]{Le81}
\bibinfo{author}{L.~Lewin}, \bibinfo{title}{Polylogarithms and associated
  functions}, \bibinfo{publisher}{North Holland}, \bibinfo{address}{New York},
  \bibinfo{year}{1981}.
\bibitem[{Abramowitz and Stegun(1970)}]{AS70}
\bibinfo{author}{M.~Abramowitz}, \bibinfo{author}{I.~A. Stegun},
  \bibinfo{title}{Handbook of Mathematical Functions},
  \bibinfo{publisher}{Dover}, \bibinfo{address}{New York},
  \bibinfo{edition}{ninth} edition, \bibinfo{year}{1970}.
\bibitem[{Mussardo(2010)}]{Mu10}
\bibinfo{author}{G.~Mussardo}, \bibinfo{title}{Statistical Field Theory: an
  Introduction to Exactly Solved Models in Statistical Physics},
  \bibinfo{publisher}{Oxford University Press}, \bibinfo{address}{Oxford},
  \bibinfo{year}{2010}.
\bibitem[{Erd{\'e}lyi(1956)}]{Er56}
\bibinfo{author}{A.~Erd{\'e}lyi}, \bibinfo{title}{Asymptotic Expansions},
  \bibinfo{publisher}{Dover}, \bibinfo{address}{New York},
  \bibinfo{year}{1956}.
\bibitem[{Baxter(1982)}]{Bax82}
\bibinfo{author}{R.~J. Baxter}, \bibinfo{title}{Exactly Solved Models in
  Statistical Mechanics}, \bibinfo{publisher}{Academic Press},
  \bibinfo{address}{London}, \bibinfo{year}{1982}.
\bibitem[{Banerjee and Basu-Mallick(2011)}]{BB11}
\bibinfo{author}{P.~Banerjee}, \bibinfo{author}{B.~Basu-Mallick},
  \bibinfo{title}{Level density distribution for one-dimensional vertex models
  related to {H}aldane--{S}hastry like spin chains}, \bibinfo{year}{2011}.
  \bibinfo{note}{ArXiv:1111.4376v2 [cond-mat.stat-mech]}.

\end{thebibliography}

\end{document}